\documentclass[a4paper,12pt]{article}

\usepackage{amsfonts,amssymb,amsthm,latexsym,graphicx,color,layout,cite}
\usepackage[english]{babel}
\usepackage[nointlimits]{amsmath}

\renewcommand{\subsection}[1]{ 
\refstepcounter{subsection}
\addcontentsline{toc}{subsection} 
      {\normalsize\normalfont\textit{\thesubsection. #1}} 
\medskip\medskip\noindent 
\normalsize\normalfont
\textit{\thesubsection. #1}\nopagebreak \par\smallskip\nopagebreak} 
\def\thesubsection{{\normalsize
{\arabic{section}.\arabic{subsection}}}} 

\numberwithin{equation}{section}
\numberwithin{theorem}{section}

\newcommand{\mc}[1]{{\mathcal #1}}
\newcommand{\mf}[1]{{\mathfrak #1}}

\newcommand{\bb}[1]{{\mathbb #1}}

\renewcommand{\epsilon}{\varepsilon}

\definecolor{light}{gray}{.9}

\begin{document}

\begin{titlepage}
\par\vskip 1cm\vskip 2em

\begin{center}
{\LARGE  \textbf{Stochastic interacting particle systems \\~\\ out of equilibrium} }

\par
\vskip 2.5em \lineskip .5em

{\large
\begin{tabular}[t]{c}
$\mbox{L. Bertini}^{1}\!\!\phantom{m}\mbox{ A. De Sole}^{2}
\!\phantom{m}\mbox{D. Gabrielli}^{3}\!\phantom{m}
\mbox{G. Jona--Lasinio}^{4}\!\phantom{m}\mbox{C. Landim}^{5}$ 
\\
\end{tabular}
\par
}
\medskip
{\small
\begin{tabular}[t]{ll}
{\bf 1} & 
Dipartimento di Matematica, Universit\`a di Roma La Sapienza\\
&  P.le A.\ Moro 2, 00185 Roma, Italy\\
&  E--mail: {\tt lorenzo@carpenter.mat.uniroma1.it}\\
{\bf 2} & Dipartimento di Matematica, Universit\`a di Roma La Sapienza, Roma, Italy\\
&  Mathematics Department, Harvard University, Cambridge MA, USA\\
&  E--mail: {\tt desole@math.harvard.edu}\\
{\bf 3} & Dipartimento di Matematica, Universit\`a dell'Aquila\\
&  67100 Coppito, L'Aquila, Italy \\
&  E--mail: {\tt gabriell@univaq.it}\\
{\bf 4} & Dipartimento di Fisica and INFN, Universit\`a di Roma La Sapienza\\
&  P.le A.\ Moro 2, 00185 Roma, Italy\\
&  E--mail: {\tt gianni.jona@roma1.infn.it}\\ 
{\bf 5}& IMPA, Estrada Dona Castorina 110, J. Botanico, 22460 Rio
de Janeiro, Brazil\\
& CNRS UPRES--A 6085, Universit\'e de Rouen,
76128 Mont--Saint--Aignan Cedex, France \\
& E--mail: {\tt landim@impa.br}\\
\end{tabular}
}
\bigskip
\end{center}

\vskip 1 em
\begin{abstract}
\noindent
This paper provides an introduction to some stochastic models of
lattice gases out of equilibrium and a discussion of results of
various kinds obtained in recent years. Although these models are
different in their microscopic features, a unified picture is emerging
at the macroscopic level, applicable, in our view, to real phenomena
where diffusion is the dominating physical mechanism. We rely mainly
on an approach developed by the authors based on the study of
dynamical large fluctuations in stationary states of open systems. The
outcome of this approach is a theory connecting the non equilibrium
thermodynamics to the transport coefficients via a variational
principle. This leads ultimately to a functional derivative equation
of Hamilton-Jacobi type for the non equilibrium free energy in which
local thermodynamic variables are the independent arguments.  In the
first part of the paper we give a detailed introduction to the
microscopic dynamics considered, while the second part, devoted to the
macroscopic properties, illustrates many consequences of the
Hamilton-Jacobi equation. In both parts several novelties are
included.
\end{abstract}

\vfill


\end{titlepage}
\vfill\eject

\section{Introduction}
\label{s:1}

Models have played a fundamental role in equilibrium statistical
mechanics.  The Ising model provided the first proof that statistical
mechanics can explain the existence of phase transitions and was a
main guide in the study of critical behavior. A reason for this
effectiveness is the circumstance that the macroscopic behavior is, to
a considerable extent, independent of the microscopic details. Hence
different systems exhibit qualitatively the same phenomenology at
large scales.

Out of equilibrium the situation is more complex. First the variety of
non equilibrium phenomena one can conceive makes it more difficult to
define general classes of phenomena for which a unified study is
possible.  Furthermore the details of the microscopic dynamics play a
far greater role than in equilibrium.  Since the first attempts to
construct a non equilibrium thermodynamics, a guiding idea has been
that of local equilibrium.  This means the following. Locally on the
macroscopic scale it is possible to define thermodynamic variables
like density, temperature, chemical potentials... which vary smoothly
on the same scale. Microscopically this implies that the system
reaches local equilibrium in a time which is short compared to the
times typical of macroscopic evolutions as described for example by
hydrodynamic equations. So what characterizes situations in which this
description applies is a separation of scales both in space and
time. There are important cases however where local equilibrium
apparently fails like aging phenomena in disordered systems due to
insufficient ergodicity.

The simplest non equilibrium states one can imagine are stationary
states of systems in contact with different reservoirs and/or under
the action of external fields. In such cases, to the contrary of
equilibrium, there are currents (electrical, heat, matter of various
chemical constitutions ...) through the system whose macroscopic
behavior is encoded in transport coefficients like the diffusion
coefficient, the conductivity or the mobility.  The ideal would be to
approach the study of these states starting from a microscopic model
of atoms interacting with realistic forces and evolving with Newtonian
dynamics. This is beyond the reach of present day mathematical tools
and much simpler models have to be adopted in the reasonable hope that
some essential features are adequately captured.

In the last decades stochastic lattice gases have provided a very useful
laboratory for  studying properties of stationary non equilibrium states
(SNS). Besides many interesting results specific to the different
models considered, the following features of general significance
have emerged.

\begin{enumerate}
\item
\label{i:1}{Local equilibrium and hydrodynamic equations have been
  derived rigorously from the microscopic dynamics for a wide class of
  stochastic models.}
\item
\label{i:2}
{A definition of non equilibrium thermodynamic functionals has
emerged via a theory of dynamic large fluctuations, moreover a general 
equation which they have to satisfy has been established. 
This is a time independent Hamilton-Jacobi (H-J) equation
whose independent arguments are the local thermodynamic variables
and requires as input the transport coefficients. These coefficients can
be either calculated explicitly for given models or obtained
from measurements so that H-J  can be used also as a phenomenological   
equation.
}
\item
\label{i:3}
{Non equilibrium long range correlations, which have been observed
experimentally in various types of fluids \cite{DKS}, appear to be generic
consequences of H-J.  An important connection between the 
behavior  of the mobility and the sign of these
correlations can be derived from the H-J equation.
}
\item
\label{i:4}
{An analysis of the fluctuations of the currents averaged over long
 times has revealed the possibility of different dynamical regimes, 
which are interpreted as dynamical phase transitions. 
Such phase transitions have actually been proved to exist in some models. 
This theoretical prediction should be investigated experimentally.
}
\end{enumerate}

As an overall comment we may say that the macroscopic theory obtained
so far encompasses the theory developed long ago by Onsager
\cite{ONS1} and then by Onsager-Machlup \cite{OMA} for states close to
equilibrium and we believe to be applicable in general to states where
diffusion is the dominant dynamical mechanism.

\medskip
The present paper intends to provide a unified introduction to models
studied intensively in the last decade for which several results have
been obtained.  We shall concentrate on time independent properties
that is mainly on the topics~\ref{i:2} and \ref{i:3} mentioned above.
Our treatment is based on an approach developed by the authors in
\cite{BDGJL1,BDGJL2,BDGJL3,BDGJL4}.
For item~\ref{i:1} we refer to \cite{KL,S,VJ} for the case of periodic
boundary conditions and to \cite{els1,els2} for open systems. 
For item~\ref{i:4}, which we do not discuss here, see 
\cite{BDGJL5,BDGJL6,BDGJL7,BDGJL8,BD1,BD2,BD3}. 
For recent overviews on nonequilibrium phenomena see also \cite{HS,MKS}.

There are two parts in the article according to the natural separation
into microscopic and macroscopic properties.  As a rule we do not
include proofs except for statements which require a short argument.
Most of the results discussed here are in published articles to which
we shall refer for the details. We do outline however the following
new results which will be the subject of forthcoming papers.  In
Section~\ref{s:kmp} we consider the KMP process \cite{kmp}. We give an
explicit representation of the invariant measure of this process in
the case of a single oscillator and we compute exactly the two point
correlations for the general case.  In Section~\ref{s:mgf} we show,
for a particular class of one-dimensional models, that nonequilibrium
long range correlations are positive if the mobility is convex and
negative if it is concave. The general case is discussed in
\cite{BDGJL10}. In Section~\ref{s:mwas} we show that for \emph{any}
weakly asymmetric model with periodic boundary conditions the
nonequilibrium free energy does not depend on the external field, so
that it coincides with the equilibrium one and there are no long range
correlations. In Section~\ref{s:mtasep} we consider the
one-dimensional boundary driven totally asymmetric exclusion process. 
For a particular choice of the parameters, 
starting from the results in \cite{DuSch} we obtained a new 
variational representation of the nonequilibrium free energy. 
In particular, while the representation in \cite{DLS3} requires the 
maximization of a trial functional, we show how it 
can be formulated as a minimization problem.

\section{Nonreversible microscopic models}
\label{s:2}

Stochastic lattice gases are -- loosely speaking -- a collection of
random walks moving in the lattice and interacting with each
other. These ``particles'' are to be considered
indistinguishable. Accordingly, the microscopic state is specified by
giving the occupation number in each site of the lattice.  The effect
of the interaction is that the \emph{jump rates} depend on the local
configuration of the particles. For non-isolated systems we model the
effect of the reservoirs by adding creation/annihilation of particles
at the boundary. The effect of an external field is modeled by
perturbing the rates and giving a net drift toward a specified
direction.

In Section~\ref{s:2.1} we give the precise definition of non
equilibrium stochastic lattice gases. Some special models, the
\emph{zero range process} and the \emph{exclusion process},
are discussed in Sections~\ref{s:2.3} and \ref{s:2.2.5}. For the zero
range process the invariant measure is always product and can be
computed explicitly. On the other hand, the boundary driven exclusion
process carries long range correlations, which can be computed
explicitly in the one dimensional case.  In Section~\ref{s:kmp} we
recall the definition of the KMP process \cite{kmp} and we 
compare it with the exclusion process.  In
Section~\ref{s:2.2} we consider \emph{gradient} lattice
gases with periodic boundary conditions; the peculiarity of such
models is that the invariant measure does not depend on the applied
external field.  In Section~\ref{s:2.4} we consider the \emph{Glauber
  + Kawasaki} model, in which a reaction term allowing
creation/annihilation of particles in the bulk is added. We discuss
under which conditions on the reaction rates it is 
reversible.  Finally, in Section~\ref{s:tasep} we consider the
boundary driven totally asymmetric exclusion process and we recall the
representation of the invariant measure obtained in
\cite{DuSch}. This representation suggests a new variational expression
for the nonequilibrium free energy that will be discussed in
Section~\ref{s:mtasep}.

\subsection{Stochastic lattice gases}
\label{s:2.1}

As basic microscopic model we consider a stochastic lattice gas in
a finite domain, with an external field, and either with periodic
boundary conditions or with particle reservoirs at the boundary.  The
process can be informally described as follows.  At each site,
independently from the others, particles wait exponential times at the
end of which one of them jumps to a neighboring site. 
In the case of particle reservoirs, in addition to this dynamics, 
we have creation and annihilation of particles, at exponential times, at
the boundary.  
To define formally the microscopic dynamics,
recall that a continuous time Markov chain $\omega_t$ on some state space $\Omega$
can be described in term of its infinitesimal \emph{generator} $L$ defined
as follows. Let $f:\Omega\to \bb R$ be an observable, then   
\begin{equation}
\bb E \big( f (\omega_{t+h}) \big| \omega_t \big) = (L f)(\omega_t) \, h
+ o(h)
\end{equation}
where $\bb E(\,|\,)$ is the conditional expectation,
so that the \emph{expected} infinitesimal increment of $f(\omega_t)$ is 
$(Lf)(\omega_t)\, dt$. The transition probability of the Markov process
$\omega_t$ is then given by the kernel of the 
semigroup generated by $L$, i.e.\ 
\begin{equation}
p_t(\omega,\omega') = e^{t L} (\omega,\omega')
\end{equation}

\smallskip
Let $\Lambda$ be the $d$-dimensional torus of side length one,
i.e.\ $(\bb R / \bb Z)^d$, respectively a smooth domain in $\bb R^d$,
and, given an integer $N> 1$, set $\Lambda_N := (\bb Z / N \bb Z)^d$,
respectively $\Lambda_N:= (N\Lambda) \cap \bb Z^d$.  The configuration
space is $X^{\Lambda_N}$, where $X$ is a subset of $\bb N$,
e.g.\ $X=\{0,1\}$ when an exclusion principle is imposed and $X=\bb N$
when there is no limitation on the number of particles.  The number of
particles at the site $x\in\Lambda_N$ is denoted by $\eta_x \in X$ and
the whole configuration by $\eta\in X^{\Lambda_N}$.  The microscopic
dynamics is then specified by a continuous time Markov chain on the
state space $X^{\Lambda_N}$ with infinitesimal generator given by $L_N
= L_{0,N}$, resp.\ $L_N = \big[ L_{0,N} + L_{b,N} \big]$, if $\Lambda$
is the torus, resp.\ a smooth domain in $\bb R^d$, where, for
functions $f:X^{\Lambda_N}\to \bb R$,
\begin{eqnarray} 
\label{gen1}
L_{0,N} f(\eta)  &= & 
\frac 12 \sum_{\substack{x,y\in\Lambda_N\\  |x-y| =1}} c_{x,y}(\eta)
\big[ f(\sigma^{x,y}\eta) - f(\eta) \big] \; 
\\
\label{gen2}
L_{b,N} f(\eta) &=& 
\frac 12
\sum_{\substack{x\in\Lambda_N, y\not\in\Lambda_N\\ |x-y|=1}}
\Big\{ c_{x,y}(\eta) \big[ f(\sigma^{x,y}\eta) - f(\eta) \big] +
c_{y,x}(\eta) \big[ f(\sigma^{y,x}\eta) - f(\eta) \big] \Big\}\qquad  
\end{eqnarray} 
Here $|x|$ stands for the usual Euclidean norm.  For
$x,y\in\Lambda_N$, $\sigma^{x,y}\eta$ is the configuration obtained
from $\eta$ by moving a particle from $x$ to $y$, i.e.
\begin{equation*}
\big( \sigma^{x,y}\eta \big)_z = \left\{
\begin{array}{lcl} \eta_z & \textrm{ if } & z\neq x,y \\ \eta_y +1 &
\textrm{ if } & z= y \\ \eta_x -1 & \textrm{ if } & z= x\;
\end{array} \right.  
\end{equation*} 
and similarly, if $x\in\Lambda_N$, $y\not\in\Lambda_N$, then
$\sigma^{y,x}\eta$ is obtained from $\eta$ by creating a particle at
$x$, while $\sigma^{x,y}\eta$ is obtained by annihilating a particle
at $x$.  Therefore for $x,y\in\Lambda_N$, $c_{x,y}$ is the rate at
which a particle at $x$ jumps to $y$. 
We assume that $c_{x,y}(\eta)=0$ if $\sigma^{x,y}\eta\not\in
X^{\Lambda_N}$ so that $L_{0,N}$ and $L_{b,N}$ are well defined linear
operators on the set of functions $f:X^{\Lambda_N}\to \bb R$.  The
generator $L_{0,N}$ describes the bulk dynamics which preserves the
total number of particles whereas $L_{b,N}$ models the particle
reservoirs at the boundary of $\Lambda_N$.

We assume that the bulk rates $c_{x,y}$, $x,y\in \Lambda_N$, are
obtained starting from reversible rates $c^0_{x,y}$ satisfying the
\emph{detailed balance} with respect to a Gibbs measure defined
by a Hamiltonian $\mc H$, and perturbing them with an external field
$F$.  Likewise, in the case of particle reservoirs, we assume that the
boundary rates $c_{x,y}$, $c_{y,x}$, $x\in\Lambda_N$,
$y\not\in\Lambda_N$, are obtained from rates $c^0_{x,y},\,c^0_{y,x}$
satisfying the local detailed balance with respect to $\mc H$ and in
presence of a chemical potential $\lambda_0$, and again perturbed by
the external field $F$.  Our analysis is restricted to the high
temperature phase, in particular we shall assume that the correlations
in the Gibbs measure decay exponentially.

The above conditions are met by the following formal definitions.
Consider jump rates $c^0_{x,y}$ satisfying the detailed balance with
respect to the Gibbs measure associated to the Hamiltonian $\mc H:\,
X^{\Lambda_N}\to\bb R$ with free boundary conditions. For the bulk
rates this means
\begin{equation}
\label{c0}
c_{x,y}^0(\eta) \;=\;
\exp\big\{ -\big[\mc H(\sigma^{x,y}\eta) - \mc H(\eta) \big]\big\}
\, c_{y,x}^0 (\sigma^{x,y} \eta)\,,
\quad x,y\in\Lambda_N
\end{equation}
Note that we included the inverse temperature in $\mc H$.  As before
if $\sigma^{x,y}\eta\not\in X^{\Lambda_N}$ we assume $c^0_{x,y}(\eta)
=0$.  From a mathematical point of view, the detailed balance
condition means that the generator is self-adjoint w.r.t.\ the Gibbs
measure $\mu(\eta) \propto e^{-\mc H (\eta) }$; namely if we let
$L_{0,N}^0$ be the generator in \eqref{gen1} with $c$ replaced by
$c^0$, for each $f,g : X^{\Lambda_N} \to \bb R$ we have
\begin{equation}
\label{sadj}
\langle f,L_{0,N}^0 g\rangle_{\mu} := 
\sum_{\eta}  \mu(\eta) \, f(\eta) \, L_{0,N}^0 g(\eta) 
= \langle L_{0,N}^0f, g\rangle_{\mu} 
\end{equation}

Let  $\overline{\Lambda}_N:= \{ x\in\bb Z^d\,|\: \exists~y\in \Lambda_N
\textrm{ with } |x-y| \le 1\}$ be the $1$-neighborhood of $\Lambda_N$. 
When $\Lambda_N$ is the discrete torus we agree that 
$\overline{\Lambda}_N={\Lambda}_N$. We also let 
$\partial \Lambda_N := \overline{\Lambda}_N\setminus {\Lambda}_N$. 
In the case when $\Lambda$ is not the torus, 
the boundary dynamics with no external field is specified as
follows. Denote by  $\lambda_0 : \partial{\Lambda}_N  \to \bb R$ the chemical
potential of the reservoirs. If $x\in\Lambda_N$, $y\not\in \Lambda_N$ 
the detailed balance condition \eqref{c0} is modified by adding the 
chemical potential $\lambda_0$:
\begin{equation}
\label{c0bound}
c_{x,y}^0(\eta) \;=\;
\exp \big\{ -\big[ \mc H(\sigma^{x,y}\eta) - \mc H(\eta)\big] 
- \lambda_0(y) \big\}
\, c_{y,x}^0 (\sigma^{x,y} \eta)\,,
\quad x\in\Lambda_N,\, y\not\in \Lambda_N\;
\end{equation}

We denote by $\mc B (\Lambda_N) :=\{ (x,y)\,|\:  x,y\in\overline{\Lambda}_N\,,\:
\{x,y\} \cap \Lambda_N \neq\emptyset \,,\; |x-y|=1\}$ 
the collections of ordered bonds intersecting  $\Lambda_N$. 
A \emph{discrete vector field} is 
is then defined as a real function  $F : \mc{B}(\Lambda_N) \to\bb R$ 
satisfying $F(x,y) = - F(y,x)$ for any $(x,y)\in \mc B (\Lambda_N)$.  
An \emph{asymmetric lattice gas} is defined by the jump rates 
\begin{equation}
\label{ratas}
c_{x,y}(\eta) \;:=\; e^{F(x,y)} 
\, c_{x,y}^0(\eta) 
\end{equation}
where $c^0$ are the unperturbed rates and $F$ is a discrete vector field.

The case of \emph{weakly asymmetric models} is obtained by choosing 
\begin{equation}
\label{wasym}
F(x,y) \equiv F_N (x,y) = E \Big( \frac{x+y}{2N} \Big) \cdot \frac{y-x}{N}
\end{equation}
where $E: \Lambda \to \bb R^d$ is a smooth vector field and $\cdot$
denotes the inner product in $\bb R^d$. 
Namely, for $N$ large, by expanding the exponential, particles
at site $x$ feel a drift $N^{-1} E(x/N)$.

\smallskip
We can rewrite the full generator $L_N$, using the notation introduced above,
as follows
\begin{equation}
\label{genmc}
L_{N} f(\eta) 
= \frac {1}2 \sum_{(x,y) \in \mc B(\Lambda_N) }
c_{x,y}(\eta) \big[ f(\sigma^{x,y}\eta) - f(\eta) \big] 
\end{equation}
Fix an initial condition $\eta\in X^{\Lambda_N}$. The trajectory of
the Markov process $\eta(t)$, $t\ge 0$, is then an element on the path
space $D\big(\bb R_+;X^{\Lambda_N}\big)$, which consists of piecewise
constant paths with values in $X^{\Lambda_N}$. We shall denote by
${\bb P}_{\eta}^N$ the probability measure on $D\big(\bb
R_+;X^{\Lambda_N}\big)$ corresponding to the distribution of the
process $\eta(t)$, $t\ge 0$, with initial condition $\eta$.
It is related to the generator $L_N$ by
${\bb P}_\eta^N(\eta(t)=\eta^\prime) = e^{tL_N}(\eta,\eta^\prime)$.

A probability measure $\mu_N$ on $X^{\Lambda_N}$ is an
\emph{invariant} measure for the process $\eta(t)$ if
\begin{equation}
\label{muinvpt}
\sum_{\eta\in X^{\Lambda_N}} \mu_N(\eta) \, e^{t L_N } (\eta,\eta') =
\mu_N (\eta')
\end{equation}
namely, if we distribute the initial condition $\eta$ according to
$\mu_N$, then the distribution of $\eta(t)$ is $\mu_N$ for any $t\ge
0$. According to general results on Markov processes, if the process
is \emph{irreducible}, i.e.\ there is a strictly positive probability
to go from any state to any other, then the invariant measure is
unique and it encodes the long time behavior of the system. More
precisely, starting from any configuration $\eta$ the distribution of
$\eta(t)$ converges to $\mu_N$ as $t\to \infty$.  In the case of a
stochastic lattice gas with particle reservoirs, if $c_{x,y}(\eta)>0$
for any $\eta\in X^{\Lambda_N}$ and any $(x,y)\in\mc B(\Lambda_N)$
such that $\sigma^{x,y}\eta\in X^{\Lambda_N}$, then the process is
irreducible and there exists a unique invariant measure. On the other
hand, if $\Lambda_N$ is the discrete torus, the total number of
particles $\sum_{x\in\Lambda_N} \eta_x$ is conserved and there exists
a one parameter family of invariant measures.  Since in general the
transition probability cannot be expressed in a closed form, condition
\eqref{muinvpt} is not convenient. However, it is easy to obtain a
necessary and sufficient infinitesimal condition for a measure to be
invariant. The measure $\mu_N$ is invariant for the process generated
by $L_N$ if and only if, for any $f:X^{\Lambda_N}\to \bb R$, we have
\begin{equation}
\label{muinv}
\mu_N \big( L_{N} f  \big) =0 
\end{equation}
where hereafter for a measure $\mu$ and an observable $f$ we denote by
$\mu(f)$ the expectation of $f$ with respect to $\mu$.

If the generator $L_N$ satisfies the detailed balance condition with
respect to some measure $\mu_N$, namely
\begin{equation}\label{rev} 
\mu_N(gL_N f) = \mu_N(f L_N g)
\qquad\qquad \forall \:f,g \,: X^{\Lambda_N} \to \bb R
\end{equation}
then $\mu_N$ is necessarily an invariant measure. In such a case the
process is said to be \emph{reversible}. This terminology is due to
the following fact.  Let $\bb P^N_{\mu_N}$ the stationary process,
i.e.\ the distribution on the path space induced by the Markov process
with initial condition distributed according to the invariant measure
$\mu_N$.  Since $\mu_N$ is invariant, the measure $\bb P^N_{\mu_N}$ is
invariant with respect to time shifts. We can thus regard $\bb
P^N_{\mu_N}$ as a measure on paths defined also for $t\le 0$, i.e.\ as
a probability on $D(\bb R; X^{\Lambda_N})$. This probability is
invariant under time reversal if and only if the measure $\mu_N$ is
reversible, i.e.\ \eqref{rev} holds.  More generally, if we denote by
$\vartheta$ the time reversal, i.e.\ $(\vartheta \eta) \, (t) :=
\eta(-t)$, we have that $\bb P^N_{\mu_N} \circ \vartheta$
is the stationary process with generator $L_N^*$, the adjoint to $L_N$
in $L_2(d\mu_N)$.  In particular, if \eqref{rev} holds, we have $\bb
P^N_{\mu_N}\circ \vartheta = \bb P^N_{\mu_N}$.

When there is a unique invariant measure which is not reversible, we
say the corresponding process is \emph{non-reversible}.  As we shall
discuss, nonequilibrium models are necessarily non-reversible, while
there exist non-reversible processes describing equilibrium phenomena,
see e.g. \cite{BJ,GJL1,GJL2,GJLV}.  The main topic that we shall
discuss is the asymptotic behavior, as $N$ diverges, of the invariant
measure $\mu_N$ for specific classes of non-reversible models.

Conditions \eqref{c0} and \eqref{c0bound} are called \emph{local}
detailed balance for the following reason. If the chemical potential
$\lambda_0$ is constant, it is easy to show that the Gibbs measure
$\mu(\eta)\propto e^{-\mc H(\eta)+\lambda_0\sum_z \eta_z}$ is
reversible with respect to the generator $L_N^0$, that is
\eqref{genmc} with rates $c^0_{x,y}$.  On the other hand, if
$\lambda_0$ is not constant, the boundary dynamics forces a current in
the system which becomes non reversible.

We discuss next the effect of the external field.  A very
particular choice of $F$ is that of a discrete \emph{gradient} vector
field, such that
\begin{equation}
  \label{grad}
  F(x,y)=\frac 12 \big[ \lambda(y)-\lambda(x) \big]
\end{equation}
for some function $\lambda: \overline{\Lambda}_N\to \bb R$.
If we further assume that $\lambda(y)=\lambda_0(y)$ for
$y\in\partial\Lambda_N$, recall that $\lambda_0$ is the chemical potential
of the boundary reservoirs, then it is easily shown that the generator
$L_N$ in \eqref{genmc} is reversible w.r.t.\ the measure 
\begin{equation}
\label{gibbsl}
\mu^{\lambda}_N(\eta)=\frac{1}{Z_N^\lambda}
\exp\Big\{- \mathcal{H}(\eta)+\sum_{x\in \Lambda_N}\lambda(x)\,\eta_x \Big\}
\end{equation}
where $Z_N^\lambda$ is the appropriate normalization constant.
In this situation the reversibility of the process is due to the fact that
the driving from the reservoirs and the external field compensate. 

\bigskip
\noindent
\emph{An equivalence principle.}\ 
We can interpret the above result from two different perspectives, getting the answer to two
opposite questions.

Consider a model with a given chemical potential $\lambda_0$ at the boundary.
We ask if we can find an external field $F$ which compensates the driving from the boundary, 
namely such that the corresponding stationary measure $\mu_N$ is reversible.
The answer to this question is certainly yes. In fact, from the above consideration,
we have a whole family of external fields that fulfill this condition: take a gradient vector field
$F$ as in \eqref{grad} with $\lambda$ such that $\lambda(y)=\lambda_0(y)$ 
for $y\in\partial\Lambda_N$.
In this case the corresponding stationary measure happens to be 
the Gibbs measure \eqref{gibbsl}.

Conversely, suppose that we have an asymmetric model with a given external field $F$.
We ask if we can find a chemical potential $\lambda_0:\,\partial\Lambda_N\to\bb R$
such that the model is reversible.
We can immediately answer affirmatively this question if we know that the external field $F$
is gradient, i.e. \eqref{grad} holds for some $\lambda:\bar\Lambda_N\to \bb R$. 
In this case we can just fix, up to an overall additive constant, 
$\lambda_0(y)=\lambda(y),\,y\in\partial\Lambda_N$.

\subsection{Zero range process}
\label{s:2.3}

The so-called \emph{zero range process} is a special case of the
lattice gases introduced in Section~\ref{s:2.1}. In each site any
number of particles is allowed so that $X=\bb N$ and the bulk symmetric jump
rates are 
\begin{equation}
\label{rzr} 
c^0_{x,y}(\eta) = g(\eta_x)\,,\qquad x,y\in\Lambda_N
,\, |x-y|=1
\end{equation}
where $g:\bb N \to \bb R_+$ is a function satisfying $g(0)=0$ with at
most linear growth. In other words, the jump rate from $x$ to $y$
depends only on the occupation number at $x$; this explains the name
of the model.  Given $\lambda_0:\partial \Lambda_N\to \bb R$, we
choose the boundary rates as
\begin{equation}
\label{rzrbound} 
c^0_{x,y}(\eta) = g(\eta_x) \,\,, \quad
c^0_{y,x}(\eta) = e^{\lambda_0(y)}\,\,,\quad
x\in\Lambda_N,\,y\in\partial\Lambda_N ,\,|x-y|=1
\end{equation}
It is not hard to check that the detailed balance conditions
\eqref{c0} and \eqref{c0bound} are satisfied with the Hamiltonian
$$
\mc H(\eta) = \sum_{x\in\Lambda_N} \log\big(g(\eta_x)!\big)
$$
where, by definition, $g(0)!=1$ and $g(k)!=g(1)\cdots g(k)$ for $k\ge1$.
The particular case in which $g$ is the identity, i.e.\ $g(k) =k$,
$k\in\bb N$, corresponds to independent random walks described in terms
of the occupation variables $\eta\in\bb N^{\Lambda_N}$.   

A peculiar feature of this model is that its invariant measure, both
with particle reservoirs and external field, is always product. 
Consider this model with external field $F$.
Denote by $\psi : \overline{\Lambda}_N \to \bb R_+$ 
the solution to 
\begin{equation}\label{laminv}
\left\{
\begin{array}{ll}
{\displaystyle 
\sum_{\substack{y\in\overline{\Lambda}_N \\ |x-y|=1}} 
[ e^{F(y,x)} \psi (y) - e^{F(x,y)}\psi (x)] = 0\,,
} & x\in\Lambda_N 
\\
{\displaystyle 
\psi (x) = \exp\{\lambda_0(x)\}\,, }
& x\in \partial{\Lambda}_N
\end{array}
\right.
\end{equation}
The invariant measure of the zero range process with external field
$F$ and boundary chemical potential $\lambda_0$ is then the
grand-canonical product measure $\mu_N=\prod_{x\in\Lambda_N}
\mu_{x}$ with marginal distributions
\begin{equation}
\mu_{x}(\eta_x = k) = \frac {1}{Z(\psi(x))} \; 
{\frac {\psi(x)^k}{g(k)!}} 
\label{INV}
\end{equation}
where 
\begin{equation}
\label{Z=}
Z(\varphi) =  \sum_{k=0}^{\infty}{\frac {\varphi^k}{g(k)!}}
\end{equation}
is the normalizing constant.
This can be verified by showing that \eqref{muinv} holds.
If $\Lambda_N$ is the discrete torus and $F$ vanishes,  any constant $\psi$ solves
\eqref{laminv}, the corresponding invariant measures are thus the
grand-canonical measures with arbitrary chemical potential. 
Since the invariant measure is always product,
the zero range process never exhibits long range correlations.

\subsection{Exclusion process}
\label{s:2.2.5}

The exclusion process is a much studied stochastic lattice gas. 
In this model an exclusion principle is imposed. 
In each site $x\in\Lambda_N$ at most one particle is allowed 
so that $X=\{0,1\}$ and there is no other interaction.
The symmetrical bulk rates are defined by 
\begin{equation}
\label{rsep}
c^0_{x,y}(\eta) = \eta_x (1-\eta_y)\,,\,\,x,y\in\Lambda_N,\,|x-y|=1
\end{equation}
namely a particle at $x$ jumps to a nearest neighbor site $y$ with
rate $1/2$ if that site is empty. 
Then \eqref{c0} holds with $\mc H=0$.
Note that the rates \eqref{rsep} satisfy  
the constraint $c^0_{x,y}(\eta) =0$ if $\sigma^{x,y} \eta\not\in
\{0,1\}^{\Lambda_N}$. 
Given a chemical potential $\lambda_0:\,\partial\Lambda_N\to\bb R$,
the local detailed balance condition \eqref{c0bound}
is met by choosing  the boundary rates as 
\begin{equation}\label{br}
c^0_{x,y}(\eta) = \eta_x K(y)\,,\,\,
c^0_{y,x}(\eta) = (1-\eta_x) K(y) e^{\lambda_0(y)}\,,\,\,
x\in\Lambda_N,\,y\in\partial\Lambda_N,\,|x-y|=1
\end{equation}
for some $K:\,\partial\Lambda_N\to\bb R_+$.

We first discuss this symmetric case. In the case of periodic boundary
conditions there is a one parameter family of invariant measures which
are the Bernoulli measures with an arbitrary parameter. Since the
total number of particles is conserved, given $k>0$, we can consider
the process on the set $\Sigma_{N,k} =
\{\eta\in \{0,1\}^{\Lambda_N} \, :\, \sum_{x\in\Lambda_N} \eta_x =k\}$.
In this set the process is irreducible and the unique invariant
measure is the uniform measure on $\Sigma_{N,k}$ which is the
\emph{canonical} ensemble associated to the Bernoulli measures. 
In the case with particles reservoirs, if the chemical potential
$\lambda_0$ is constant then the unique invariant measure is the
Bernoulli measure with parameter $\bar\rho = e^{\lambda_0}/(1+e^{\lambda_0})$,
i.e.\ $\mu_N(\eta)=e^{\lambda_0\sum_{x\in\Lambda_N}\eta_x}/(1+e^{\lambda_0})^{|\Lambda_N|}$.
In both these situations the process is reversible.

\bigskip
\noindent\emph{One dimensional boundary driven exclusion process.} \
Unlike the zero range model, if $\lambda_0$ is not constant, so that
this becomes a nonequilibrium model, the invariant measure is not
a product measure and carries long range correlations. Let us discuss in more
detail the one-dimensional case. Assume that $\Lambda=(0,1)$ so that 
$\Lambda_N = \{1,\ldots,N-1\}$; we also let $\lambda_0:=\lambda_0(0)$ and
$\lambda_1:=\lambda_0(N)$ be the two chemical potentials of the reservoirs.

An old result by Kingman \cite{K} computes the marginals of the unique
invariant measure for a special choice of the injection rates.  More
precisely, in the case analyzed by Kingman the bulk rates are as in
\eqref{rsep} while the boundary rates are obtained by the following
limiting procedure.  In \eqref{br} choose $K(0)= (e^{A} + e^{-A})^{-1}
e^{-\lambda_0/2}$ and $K(N)= (e^{A} + e^{-A})^{-1} e^{-\lambda_1/2}$
for some $A\in\bb R$.  Consider then the asymmetric model with rates
$c_{x,y}$ as in \eqref{ratas} by introducing the external field $F$
given by $F(0,1)=F(N-1,N)={A}$, $F(1,0)=F(N,N-1)={-A}$, and $F(x,y)=0$
in all the remaining bonds.  Finally we take the limit $A\to\infty$
obtaining 
\begin{equation}
\label{br2}
c_{0,1}(\eta) = (1-\eta_1) \, e^{\lambda_0/2} \,,\quad
c_{N-1,N}(\eta) = \eta_{N-1} \, e^{-\lambda_1/2} \,,\qquad
c_{1,0}(\eta) =c_{N,N-1}(\eta) =0
\end{equation}
i.e.\ from the left endpoint particles enter with rate $(1/2)
e^{\lambda_0/2}$ but do not exit, while particles from the right
endpoint exit with rate $(1/2) e^{-\lambda_1/2}$ but do not enter.

By some smart duality computations, Kingman shows that, for this
particular choice of the boundary rates,  
the marginals of the invariant measure $\mu_N$ are
\begin{equation}
\mu_N\left(\eta_{x_1}=1,\cdots
,\eta_{x_m}=1\right)=\frac{(A-m-x_1)(A-m-x_2+1)\cdots
(A-1-x_m)}{(B-m)(B-m+1)\cdots (B-1)}
\end{equation}
where $1\leq x_1<x_2<\cdots <x_m\leq N-1$ are lattice sites and the
parameters $A$ and $B$ are defined as
\begin{equation}
A=N+e^{\lambda_1/2}\,; \quad
B=N-1+e^{\lambda_1/2}+e^{-{\lambda_0/2}}
\end{equation}

\bigskip
More recent work based on matrix methods, allows to get some
representation of the invariant measure in the general one-dimensional
case, see e.g.\ \cite{Li2,Sc} and references therein.

\smallskip
We consider now the one-dimensional boundary driven symmetric exclusion model 
with boundary rates as in \eqref{br} with 
$K(0)= (1+e^{\lambda_0})^{-1}$ and $K(N)= (1+e^{\lambda_1})^{-1}$. As
before $\lambda_0$ and $\lambda_1$ are the chemical potentials of the 
boundary reservoirs.
Letting $\rho_{i} = e^{\lambda_i}/(1+e^{\lambda_i})$, $i=0,1$, be the
corresponding densities, we then get 
\begin{equation*}
c_{1,0}(\eta)= (1-\rho_0) \eta_1 \,,
\;  c_{0,1}(\eta) = \rho_0 (1-\eta_1)  
\,,\;
c_{N-1,N}(\eta)=  (1-\rho_1) \eta_{N-1}\,,
\;
c_{N,N-1} =\rho_1 (1-\eta_{N-1}) 
\end{equation*}

Let $\mu_N$ be the unique invariant measure, it is not difficult to show that 
the density profile $\mu_N(\eta_x)$ is linear so that 
\begin{equation}
\label{f01bis}
\mu_N(\eta_x) = \rho_0  + \frac xN \, (\rho_1 -\rho_0)
\end{equation}
As first shown in \cite{S1}, it is also possible to obtain a closed
expression for the two-point correlations. 
For $1\le x< y\le N-1$ we have
\begin{equation}
\label{covSEP}
\mu_{N} (\eta_x; \eta_y) 
:= \mu_N \big(\eta_x \eta_y\big) - \mu_N(\eta_x) \mu_N(\eta_y) 
 = - \frac { (\rho_1 -\rho_0)^2}{N-1} \, \frac xN  \Big( 1 - \frac yN \Big)
\end{equation}
To prove this result it is enough to compute $L_N(\eta_x\eta_y)$,
i.e.\ the action of $L_N$ on the function $\eta_x\eta_y$, and
solve the equation 
\begin{equation}
\mu_N\big( L_N (\eta_x \eta_y) \big)=0
\end{equation}

Note that, if we take $x<y$ at distance $O(N)$ from the boundary, then
the covariance between $\eta_x$ and $\eta_y$ is of order $O(1/N)$.
Moreover the random variables $\eta_x$ and $\eta_y$ are negatively
correlated. This is the same qualitative behavior of the two-point
correlation for the uniform measure on $\Sigma_{N,k}$.
As we shall show below, quite the opposite behavior is found in another
model, the KMP process.

\bigskip
\noindent\emph{One dimensional periodic asymmetric exclusion process.} \
We finally discuss the case of the asymmetric exclusion process on
the discrete torus. It is defined by the jump rates 
\begin{equation}
\label{rasep}
c_{x,x+1}(\eta) = e^{F} \eta_x (1-\eta_{x+1})
\qquad 
c_{x+1,x}(\eta) = e^{-F} \eta_{x+1} (1-\eta_x)
\end{equation}
for some $F\in\bb R$ so that \eqref{ratas} holds with constant
external field $F$. 
A simple
computation shows that the Bernoulli measure $\mu_{\rho}$ with arbitrary
$\rho\in[0,1]$ is an invariant measure. Note however that for $F\neq 0$
the process is not reversible; in fact the stationary process
w.r.t.\ $\mu_\rho$ carries the mean current $\rho(1-\rho)\sinh(F)$. 
Unlike the zero range, if the external field $F$ is not constant the
invariant measures are in general not anymore product. Note however
that if $F$ is a \emph{gradient} vector field, as shown before,
the process is reversible w.r.t.\ a product measure. 
We emphasize that a constant vector field on the torus is not gradient.

\subsection{The boundary driven KMP process}
\label{s:kmp}

The Kipnis-Marchioro-Presutti (KMP) model \cite{kmp} 
describes a chain of one-dimensional harmonic oscillators 
which are mechanically uncoupled but interact stochastically as follows. 
Each pair of nearest neighbors oscillators waits an exponential 
time of rate one and then redistributes uniformly its total energy. 
The two oscillators at the end points are coupled to heat
reservoirs. 
Since the single spin space state is not discrete and the
elementary dynamics is associated to the bonds,  this model does not
really fit in the framework introduced in Section~\ref{s:2}. However
the precise definition of the model is straightforward. 
Let $\Lambda=(0,1)$ so that $\Lambda_N = N \Lambda \cap \bb Z \equiv
\{1,\ldots,N-1\}$. We denote by $\xi_x$ the energy of
the oscillator at the site $x\in\Lambda_N$, so that the state space is
$\bb R_+^{\Lambda_N}$. On it we introduce the Markov generator $L_N$
as follows. Given $(x,y)\in\mc B (\Lambda_N)$ and $p\in[0,1]$ we let 
$\xi^{(x,y) ,p}$ be the configuration obtained from $\xi$ by
moving a fraction $p$ of the total energy $\xi_x+\xi_y$
across the bond $(x,y)$ to $x$ and a fraction $1-p$ to $y$, i.e.\
\begin{equation*}
(\xi^{(x,y),p})_z:=\left\{
\begin{array}{ll}
\xi_z & \textrm{ if } \ \ z\neq x,y \\
p\, (\xi_x+\xi_{y}) & \textrm{ if  }\ \ z=x \\
(1-p)\, (\xi_x+\xi_{y}) & \textrm{ if }\ \ z=y \\
\end{array}
\right.
\end{equation*}
We then set $  L_N :=   \sum_{x=0}^{N-1} L_{x,x+1} $ 
where, for $f:\bb R_+^{\Lambda_N}\to\bb R$,  the bulk dynamics is given by 
\begin{equation*}
L_{x,x+1} f (\xi) :=
\int_0^1\!dp \:\big[ f( \xi^{(x,x+1),p} ) - f( \xi) \big]\,\,,\,\,\,\,
x=1,\dots,N-2
\end{equation*}
while  the boundary generators $L_{0,1}$ and $L_{N-1,N}$ are 
\begin{eqnarray}
\label{genb0}
\nonumber
L_{0,1} f (\xi) &:= &
\int_0^\infty\!d\xi_0\, \frac{1}{T_0} \, e^{-\frac{\xi_0}{T_0}} 
\int_0^1\!dp \:\big[ f( \xi^{(0,1),p} ) - f( \xi) \big]\\
L_{N-1,N} f (\xi) &:= &
\int_0^\infty\!d\xi_N\, \frac{1}{T_1} \, e^{-\frac{\xi_N}{T_1}} 
\int_0^1\!dp \:\big[ f( \xi^{(N-1,N),p} ) - f( \xi) \big]
\end{eqnarray} 
Namely, we suppose that there is an energy exchange across the ghost bonds
$(0,1)$ and $(N-1,N)$, and we put at the sites $0$ and $N$
oscillators whose energies are randomly chosen according to the
Gibbs distributions with temperatures $T_0$ and $T_1$. 

We emphasize that the above choice of the boundary dynamics differs
slightly from the original one in \cite{kmp}. 
Besides being more natural, this choice simplifies some microscopic
computations. 
Note that in the case $T=T_0=T_1$, namely of an equilibrium model, the
above process is reversible with respect to the Gibbs measure
\begin{equation}
\label{prodexp}
d\mu_N(\xi) = \prod_{x=1}^{N-1} \frac {1}{T}    e^{ - \xi_x/T} \,{d\xi_x}  
\end{equation}
which is just the product of exponential distributions. 

Later, in order to find a closed expression for the microscopic
two-point correlation functions, we introduce a more general class of
boundary dynamics which is obtained by replacing in \eqref{genb0} the
two exponential distribution on boundary sites $0$ and $N$ by other
two probability measures on $\bb R_+$.  Of course the macroscopic
behavior is the same for any reasonable choice of the boundary
dynamics.

\bigskip
\noindent\emph{Invariant measure for a single oscillator.} \
We consider the KMP model with a single oscillator, i.e.\ $N=2$. Even
in this case, as the system is in thermal contact with two reservoirs,
its stationary state is not trivial. We next show that the invariant
measure is a mixture of the Gibbs distributions with temperatures
between $T_0$ and $T_1$. Furthermore we compute the weight of each
distribution which turns out to be the arcsin law in the
interval $[T_0,T_1]$, here we assume $T_0\le T_1$. We emphasize that
this result depends on the specific choice of the boundary dynamics.

We claim that the invariant measure (a probability
measure on $\bb R_+$) is absolutely continuous w.r.t.\ the Lebesgue
measure $d\xi$ and its density can be expressed as
\begin{equation}
  \label{4.2}
   \frac{d\mu}{d\xi} = \int_{T_0}^{T_1} 
   \! d\varrho_{T_0,T_1} (T) \; \frac{1}{T} \, e^{ - \xi/T} 
\end{equation}
where $\varrho_{T_0,T_1}$ is the arcsin distribution in the interval
$[T_0,T_1]$, namely for $T$ in this interval we have 
\begin{equation}
  \label{4.2.5}
  d\varrho_{T_0,T_1}(T)  = \frac{1}{\pi} \, \frac{1}{\sqrt{(T_1-T)(T-T_0)}} \,dT
\end{equation}   

To show that $\mu$ in \eqref{4.2} is the invariant measure of the KMP
process with a single oscillator, we need to check that for
each smooth real function $f$ on $\bb R_+$ \eqref{muinv} holds.
By linearity and approximation by linear combinations of exponential
functions, it is enough to show that  \eqref{muinv} holds
if $f(\xi) = \exp\{-\lambda\xi\}$, $\lambda>0$. With this choice we
have 
\begin{eqnarray*}
   L f(\xi) &=&  
\int_0^\infty \!\frac{d\xi_0}{T_0} \, e^{- \xi_0/T_0}  
\int_0^\infty \!\frac{d\xi_2}{T_1} \, e^{- \xi_2/T_1}  
\int_0^1 \! dp \, \Big[ e^{-\lambda p(\xi_{0} +\xi)} + 
  e^{-\lambda p (\xi +\xi_2)} - 2 e^{-\lambda\xi} \Big]
\\
&=& \int_0^1\!dp \, 
\Big[ \frac{1}{1+\lambda p T_0} e^{-\lambda p\xi } 
+   \frac{1}{1+\lambda p T_1} 
e^{-\lambda p\xi} - 2 e^{-\lambda\xi}\Big] 
\end{eqnarray*}
If we now take the average of the above expression when $\xi$ is an
exponential random variable of parameter $T$ we get 
\begin{equation}
  \label{4.3.5}
  \int_0^\infty \!\frac{d\xi}{T}  e^{-\xi/T} \, L f(\xi)
  =  \int_0^1\!dp \, 
\Big[ \frac{1}{(1+\lambda p T_0)(1+\lambda p T) }  
+   \frac{1}{(1+\lambda p T) (1+\lambda p T_1)} 
- \frac{2}{1+\lambda T} \Big]
\end{equation}

We next note that the arcsin distribution in the interval $[T_0,T_1]$
is characterized by the following property. For each $\gamma\ge 0$ we
have 
\begin{equation}
    \label{4.4}
  \int \!d \varrho_{T_0,T_1}(T)
   \; \frac{1}{1+\gamma T} = \frac 1{\sqrt{(1+\gamma T_0) (1+\gamma T_1)}}
\end{equation}
The integral on the l.h.s\ can be in fact computed by using the
density in \eqref{4.2.5} and the residue theorem. Conversely, by
expanding the above equation in power series of $\gamma$, we get that 
the moments of $\varrho_{T_0,T_1}$ are determined.

Recalling \eqref{4.3.5}, to complete the proof of \eqref{4.2} it
remains to show that
\begin{equation*}
 \int \!d \varrho_{T_0,T_1}(T)\int_0^1\!dp \, 
 \bigg[ \frac{1}{(1+\lambda p T_0)(1+\lambda p T) }  
 +  \frac{1}{(1+\lambda p T) (1+\lambda p T_1)} 
 - \frac{2}{1+\lambda T} \bigg] = 0 
\end{equation*}
which in view of \eqref{4.4} is equivalent to 
\begin{eqnarray*}
&& \int_0^1\!dp \, 
\bigg[ \frac{1}{ (1+\lambda p T_0)^{3/2}(1+\lambda p T_1)^{1/2}}  
+   \frac{1}{(1+\lambda p T_0)^{1/2} (1+\lambda p T_1)^{3/2} } 
\bigg] 
\\
&&\qquad\qquad \vphantom{\big\{^{\Big\{ }}
= \frac{2}{ (1+\lambda T_0)^{1/2} (1+\lambda T_1)^{1/2} } 
\end{eqnarray*}
By a direct integration we get 
\begin{equation*}
   \int_0^1\!dp \,  \frac{1}{(1+\lambda p T_0)^{3/2}(1+\lambda p T_1)^{1/2}}
     = \frac{2}{\lambda (T_1-T_0)} 
     \bigg[ \frac{ \sqrt{1+\lambda T_1}}{\sqrt{1+\lambda T_0}} -1 \bigg]
\end{equation*}
and simple algebraic computations yield the result. 

It seems quite hard to obtain an analogous representation for $N\ge
3$. On the other hand we compute explicitly below the one and two
point correlation functions of $\mu_N$ for any $N\ge 2$.

\bigskip
\noindent\emph{Two point correlations.}\ 
We here consider the KMP process with boundary dynamics given by 
\begin{eqnarray*}
\nonumber
L_{0,1} f (\xi) &:= &
\int_0^\infty\!d\nu_0^N(\xi_0) \, 
\int_0^1\!dp \:\big[ f( \xi^{(0,1),p} ) - f( \xi) \big]\\
L_{N-1,N} f (\xi) &:= &
\int_0^\infty\!d\nu_1^N(\xi_N) 
\int_0^1\!dp \:\big[ f( \xi^{(N-1,N),p} ) - f( \xi) \big]
\end{eqnarray*} 
where $\nu_i^N$, $i=0,1$, are probability measures on $\bb R_+$ with mean 
$T_i$ and variance 
\begin{equation}
\label{cov01}
\nu^N_i \big( [\xi - T_i]^2 \big) = T_i^2 + \frac{(T_1-T_0)^2}{N(N+1)}
\qquad i=0,1
\end{equation}
and note that the exponential distributions chosen in \eqref{genb0}
fail to satisfy the above condition only by a term $O(N^{-2})$.

Given $T_0\le T_1$,
let $\mu_N$ be the invariant measure of the KMP process with $N-1$
oscillators
and set  $E_N(x) := \mu_N(\xi_x)$, $x=1,\cdots,N-1$, as well as 
$E_N(0):=T_0$, $E_N(N):=T_1$.  By choosing
linear functions $f$ in \eqref{muinv} and computing 
$L_N\xi_x$, $x=1,\cdots,N-1$, we get a closed equation for $E_N$ 
which yields
\begin{equation}
\label{1pointKMP}
E_N(x) = T_0 + (T_1-T_0) \frac xN
\end{equation}

Let $C_N(x,y) := \mu_N(\xi_x;\xi_y)= \mu_N(\xi_x\,\xi_y)
-E_N(x)E_N(y)$, $x,y\in\{ 1,\cdots,N-1\}$, be the two point correlation
function of $\mu_N$. 
We also set $C_N(0,0):= \nu^N_0\big([\xi-T_0]^2\big)$, 
$C_N(N,N):=\nu^N_1\big([\xi-T_1]^2\big)$, 
$C_N(0,y)=C_N(x,N):=0$ for  $1\le y \le N$,  $0\le x \le N-1$.   
By choosing quadratic functions in \eqref{muinv}, by some elementary
but tedious computations we get that $C_N(\cdot,\cdot)$ solves 
\begin{equation*}
\left\{ 
\begin{array}{l}
\big(\Delta^N_x + \Delta^N_y \big) C_N(x,y) = 0 
\qquad \qquad \qquad \qquad \;\; 
1\le x \le y \le N-1\,,\quad y-x\ge 2 \\  
\\
C_N(x,x+2)  + C_N(x-1,x+1) - \frac{10}3 \, C_N(x,x+1)   
+ \frac{1}3 \, C_N(x,x)  
\\
\quad 
+ \frac{1}3 \, C_N(x+1,x+1)   
= \frac{2}3 \, \big( \frac{T_1-T_0}N \big)^2  + \frac{2}3 \, E_N(x) \, E_N(x+1) 
\qquad 1\le x \le N-2\\ 
\\
C_N(x-1,x-1)  + C_N(x+1,x+1) +2 \, C_N(x-1,x) + 2 \, C_N(x,x+1)
\\
\qquad 
- 4 C_N(x,x)  
= -2 \big( \frac{T_1-T_0}N \big)^2  -2 \, E_N(x)^2 
\qquad \qquad \qquad \qquad \;\;
1\le x \le N-1
\end{array}
\right.
\end{equation*}
where  $\Delta^N_xf(x,y) = f(x+1,y) + f(x-1,y) -2 f(x,y)$
is the discrete Laplacian w.r.t.\ $x$ and  
$\Delta^N_y$ is the discrete Laplacian w.r.t.\ $y$.

As can be easily checked, the solution is given by
\begin{equation}
\label{covKMP}
C_N(x,y)= 
\left\{ 
\begin{array}{ll}
\frac{(T_1-T_0)^2}{N+1} \frac xN \big( 1- \frac yN\big) 
&  0\le x< y \le N \\ 
\\
E_N(x)^2 + 2 
\frac{(T_1-T_0)^2}{N+1} \frac xN \big( 1- \frac xN\big) 
+ \frac{(T_1-T_0)^2}{N(N+1)} 
&  0\le x=y \le N \\ 
\end{array}
\right.
\end{equation}

Comparing \eqref{covKMP} with \eqref{covSEP} we observe that the off
diagonal terms are essentially the same in the macroscopic limit $N\to
\infty$. We emphasize however that the sign is different: while for
the boundary driven symmetric exclusion the occupation variables
$\eta_x$, $x\in\Lambda_N$, are negatively correlated, for the KMP
process the local energies $\xi_x$, $x\in\Lambda_N$, are positively
correlated.  As we shall discuss in Section~\ref{s:mgf}, this
qualitative difference is related to the different convexity
properties of the mobilities of the two models. For the exclusion it 
is concave while it is convex for KMP.

\smallskip
We mention that an analogous computation has been recently performed
for a somewhat similar model, see \cite{KGR}.

\subsection{Gradient models with periodic boundary conditions} 
\label{s:2.2}

In this section we consider only the case of periodic boundary
conditions, namely $\Lambda_N$ is the discrete torus $(\bb Z / N \bb
Z)^d$. We also assume that the model is translationally
covariant in the sense that, for any $(x,y)\in \mc B (\Lambda_N)$,
$z\in \Lambda_N$, and $\eta\in X^{\Lambda_N}$, we have
\begin{equation}
\label{trinv}
c_{x,y}(\eta) = c_{x+z,y+z}(\tau_z \eta) 
\end{equation}
where $\tau_z$ is the space shift, i.e.\ $(\tau_z\eta)_{x} := \eta_{x-z}$.

Let the bulk rates $c^0_{x,y}$ satisfy \eqref{c0}.  The \emph{expected
instantaneous current} across the bond $(x,y)\in \mc B(\Lambda_N)$ is,
up to a factor 2,
\begin{equation*}
j^0_{x,y}(\eta):=c^0_{x,y}(\eta) - c^0_{y,x}(\eta)
\end{equation*}
The corresponding lattice gas (with no external field) 
satisfies the \emph{gradient condition} if the discrete vector field
$j^0_{x,y}(\eta)$ is gradient for any $\eta\in X^{\Lambda_N}$, namely
there exist functions $h_x:  X^{\Lambda_N} \to \bb R$, $x\in
\Lambda_N$,  such that for any  $(x,y) \in \mc B(\Lambda_N) $
\begin{equation}
\label{gradient}
j^0_{x,y}(\eta)= h_y(\eta) - h_x(\eta)
\end{equation}
The zero range process of Section~\ref{s:2.3} 
is a gradient lattice gas for any choice of the function $g$. 
Indeed, for the rates \eqref{rzr} condition \eqref{gradient} holds
with $h_x(\eta)=-g(\eta_x)$.
Also the exclusion process is gradient, \eqref{gradient} holding 
with $h_x(\eta)= -\eta_x$.

In the stochastic gases literature, see \cite{KL,S}, the gradient
condition is usually stated in a stronger form. More precisely, one
considers a translationally invariant lattice gas on the whole lattice
$\bb Z^d$ and says that the model is gradient if there exists a
function ${\tilde h}: X^{{\bb Z}^d} \to \bb R$ which is \emph{local},
i.e.\ it depends on $\eta_x$ only for a finite number of $x\in\bb
Z^d$, and such that for any $(x,y)\in\mc B (\bb Z^d)$ and $\eta \in
X^{{\bb Z}^d}$
\begin{equation}
\label{stgrad}
j^0_{x,y}(\eta) = \tilde h(\tau_y\eta) -\tilde h(\tau_x\eta)
\end{equation}
Of course \eqref{stgrad} implies \eqref{gradient} for $N$ large
enough.  Conversely, it is possible to show that if \eqref{gradient}
holds then there exists a function $\tilde h : X^{\Lambda_N} \to \bb R$
such that \eqref{stgrad} holds for any $(x,y)\in\mc B(\Lambda_N)$.

Consider now a lattice gas with constant (non zero) external field $F$.
By this we mean that $F(x,x\pm e_i)= \pm F_i$, where $e_i$, $i=1,\cdots,d$ is
the canonical basis in $\bb R^d$ and $(F_1,\dots ,F_d)$ is a vector in 
$\bb R^d$.  
As discussed in \cite{KLS}, if the bulk rates $c^0$ satisfy the
gradient condition \eqref{stgrad} then the grand-canonical Gibbs
measures $\exp\{ - \mc H(\eta) +\lambda \sum_{x\in\Lambda_N}
\eta_x\}$, $\lambda\in\bb R$, which are the invariant measures for the
system with no external field, are invariant also for the process with
external field $F$, i.e.\ with rates $c_{x,x\pm e_i}= c^0_{x,x\pm e_i} 
e^{\pm F_i}$, $i=1,\cdots, d$.  In particular this result shows that
\emph{gradient} lattice gases with constant external field and
periodic boundary conditions do not exhibit long range correlations.
In Section~\ref{s:was} we show that, from a macroscopic point of view,
\emph{any} weakly asymmetric lattice gas with periodic boundary
conditions does not have long range correlations.

We next discuss, from a microscopic point of view, 
gradient lattice gases in some more detail obtaining the above
mentioned result as a particular case. Let us consider an
asymmetric lattice gas with 
external field $F:\mc B(\Lambda_N)\to \bb R$, rates $c_{x,y}$ as in
\eqref{ratas}, and  generator given by \eqref{genmc}. 
We look for an invariant measure of the form \eqref{gibbsl}
for some $\lambda:\Lambda_N\to \mathbb R$. The condition for a 
stationary state is
\begin{equation}
  \label{sta1}
\sum_{\eta\in X^{\Lambda_N}}\mu_N^{\lambda}(\eta)
\sum_{(x,y)\in\mc B(\Lambda_N)} 
c_{x,y}(\eta)\big[f(\sigma^{x,y}\eta)-f(\eta)\big] =0\,,
\quad \forall \, f: X^{\Lambda_N} \to \bb R.
 \end{equation}
Performing some change of variables and using the conditions
\eqref{c0} and \eqref{c0bound} of local detailed balance, \eqref{sta1} 
becomes
\begin{equation}
\label{sta2}
\sum_{\eta\in X^{\Lambda_N}}f(\eta)\mu_N^{\lambda}(\eta) 
\sum_{(x,y)\in\mc B(\Lambda_N)} 
e^{-\lambda(x)}e^{F(x,y)}\left[e^{\lambda(y)}c^0_{y,x}(\eta)
-e^{\lambda (x)}c^0_{x,y}(\eta)\right] =0
\end{equation}
Let  
\begin{eqnarray*}
G^{\lambda}(x,y) &:=& e^{-\lambda(x)}e^{F(x,y)}-e^{-\lambda(y)}e^{F(y,x)} \\
j^{\lambda}_{y,x}(\eta) &:=& e^{\lambda(y)}c^0_{y,x}(\eta)    
-e^{\lambda(x)}c^0_{x,y}(\eta)
\end{eqnarray*}
Note that if $F$ is a discrete vector field, 
i.e.\ it satisfies $F(x,y)=-F(y,x)$, then $e^F$ is \emph{not} a discrete 
vector field but $G^{\lambda}$ and $j^{\lambda}$ are
indeed discrete vector fields. We then get that \eqref{sta2} is equivalent to
\begin{equation}
\label{sta3}
\sum_{(x,y) \in \mc B(\Lambda_N)}  
G^{\lambda}(x,y)\, j^{\lambda}_{y,x}(\eta)=0\,, \quad
\forall \eta\in X^{\Lambda_N}.
\end{equation}
Notice that \eqref{sta3} is an orthogonality condition. 
In general there is no solution to \eqref{sta3}; note in fact that 
it is a system of $|X|^{|\Lambda_N|}$ equations 
(corresponding to different particles configurations) 
but we have only $|\Lambda_N|$ parameters (corresponding to the chemical
potential profile $\lambda:\Lambda_N\to\bb R$). 
Non-existence of solutions to \eqref{sta3} means
that the invariant measure is not of the form \eqref{gibbsl}.
There are however few remarkable cases in which \eqref{sta3} can be
easily solved.

If the model is gradient, so that \eqref{gradient} holds, we claim
that $\mu_N^{\lambda}(\eta)$ in \eqref{gibbsl} with $\lambda \in \bb R$
constant is an invariant measure for any asymmetric lattice gas
provided the external field $F$ satisfies
\begin{equation}
\label{circulation}
\sum_{y : |x-y|=1} \Big[ e^{F(x,y)} - e^{F(y,x)} \Big]= 0
\qquad \forall\, x\in\Lambda_N
\end{equation}
that is the discrete vector field $G^\lambda$, $\lambda\in\bb R$, has
vanishing discrete divergence. 
Conversely, if we require that $\mu_N^{\lambda}(\eta)$, $\lambda \in \bb
R$, is an invariant measure for any external field satisfying
\eqref{circulation}, we get that the rates $c^0_{x,y}$ have 
to satisfy \eqref{gradient} for some functions $h_x\,: X^{\Lambda_N}
\to \bb R$, $x\in\Lambda_N$. 
The proof of both statements are accomplished by some computations
which essentially amounts to prove the Hodge theorem in a discrete
setting, see \cite{Lo}.

\medskip
\noindent
\emph{Generalized gradient models.}\ 
Consider asymmetric lattice gases with constant external fields $F$.
Some computations show that $\mu_N^\lambda$ as in \eqref{gibbsl} is
an invariant measure for any constant $\lambda\in\bb R$ 
if and only if the rates $c^0_{x,y}$ satisfy
\begin{equation}
  \label{ort}
\sum_{x\in \Lambda_N}j^0_{x,x+e_i}(\eta)=0\,, \quad \forall
\eta\in X^{\Lambda_N} \,,\forall i=1,\dots ,d.
\end{equation}
which is exactly the condition that identifies the orthogonal
complement, w.r.t.\ the inner product defined in \eqref{sta3},
of the constant vector fields.  Moreover \eqref{ort} is equivalent to
the following \emph{generalized gradient condition}. There are
function $h_{i,j}:X^{\Lambda_N}\to \bb R$, $i,j=1,\cdots,d$, such that
\begin{equation}
  \label{ggrad}
  j^0_{x,x+e_i}(\eta)=\sum_{j=1}^d 
  \big[ h_{i,j}(\tau_{x+e_j}\eta) - h_{i,j} (\tau_{x}\eta) \big]\,,
\qquad i=1,\cdots,d 
\end{equation}
We finally mention that \eqref{ggrad} is a particular case of
the condition stated in \cite[Def.~2.5]{KL}.

\medskip
To summarize the previous discussion, gradient models in the sense of
\eqref{gradient} have the property that any external field satisfying
\eqref{circulation} will not change the invariant measure, while generalized
gradient models in the sense of \eqref{ggrad} have this property only
for constant external fields.

\subsection{Glauber + Kawasaki dynamics}
\label{s:2.4}

Unlike the models discussed so far, the so-called \emph{Glauber +
Kawasaki} process is not a lattice gas in the sense that the number
of particles is not locally conserved. A reaction term allowing
creation/annihilation of particles is added in the bulk.
We consider the case with exclusion rule so that $X=\{0,1\}$ and
discuss only the one-dimensional case with periodic boundary
condition, $\Lambda_N$ a ring with $N$ sites. The generator is defined as 
\begin{equation}
\label{eqn1}
L_N f(\eta) = \frac{1}{2} 
\sum_{(x,y)\in \mc B (\Lambda_N)} 
\eta_x(1-\eta_y) 
\big[ f(\sigma^{x,y} \eta )-f(\eta) \big] 
+ \frac1{N^2} \sum_{x\in \Lambda_N} c_x(\eta) 
\big[ f(\sigma^{x} \eta)-f(\eta) \big]
\end{equation}
where $\sigma^x$ denotes the particle flip at $x$, i.e.\ 
$(\sigma^x \eta)_x = 1-\eta_x$ and  $(\sigma^x \eta)_y = \eta_y$ for 
$y\neq x$.   
The first term of the generator corresponds to the symmetric  exclusion
process while the second one involves the reaction defined by the
corresponding rates $c_x$, $x\in\Lambda_N$.
The factor $N^2$ in \eqref{eqn1} has been inserted to get, after
diffusive rescaling, a meaningful macroscopic evolution.

The first question one can ask is when there exists
a reversible measure for this process. As we shall see, this happens 
only if we impose some restrictions on the reaction rates $c_x$.
The condition of reversibility w.r.t.\ the measure $\mu_N$ is
\eqref{sadj}, which in this case, after some algebra, reads
\begin{eqnarray}
\label{eqn23}
\nonumber
& & 
\frac{1}{2}  
\sum_{\eta} 
\sum_{x\in\Lambda_N} 
g(\eta) \, f(\eta^{x,x+1}) 
\big[ \mu_N (\eta) -\mu_N(\eta^{x,x+1}) \big] \\
& & \quad + 
\frac 1{N^2} \sum_{\eta} \sum_{x\in\Lambda_N}   
g(\eta) \, f(\sigma^x \eta) 
\big[  c_x(\eta) \mu_N(\eta)- c_x(\sigma^x\eta) \mu_N(\sigma^x \eta) \big]=0
\end{eqnarray}
where $\eta^{x,x+1}$ denotes the configuration obtained from $\eta$ by
exchanging the occupation numbers in $x$ and $x+1$.
Since this equality must hold for every $g$ and $f$, this condition is 
equivalent to
\begin{equation}
\left\{
\begin{array}{ccl}
\mu_N(\eta)-\mu_N(\eta^{x,x+1}) &=& 0 \\
c_x(\eta) \mu_N(\eta) - c_x (\sigma^x\eta) \, 
\mu_N(\sigma^x \eta) &=& 0
\end{array}
\right.
\label{eqn24}
\end{equation}
for any $\eta$ and $x$. 
The first condition imposes that the measure $\mu_N$  has the form 
\begin{equation}
\label{eqn25}
\mu_N(\eta)= M_N\Big( \sum_{x\in \Lambda_N} \eta_x \Big)
\end{equation}
namely $\mu_N$ must assign an equal weight to configurations with the
same number of particles.  The second condition, with a $\mu_N$ of
this type, is a restriction on the reaction rates and on the
function $M_N$. The most general form of $c_x(\eta)$ that satisfies
this condition is
\begin{equation}
\label{eqn26}
c_x (\eta)= A_1  \, (1-\eta_x) \, h(\tau_x\eta) + A_2\, \eta_x  h(\tau_x\eta)
\end{equation}
where $A_1,A_2$ are arbitrary positive constants, and $h:
\{0,1\}^{\Lambda_N} \to \bb R_+$ is an arbitrary positive function
such that $h(\sigma^0 \eta) = h (\eta)$, i.e.\ it does not depend on
$\eta_0$.  Recall that $\tau_x$ denotes the shift by $x$.  Notice that
the rates $c_x(\eta)$ in \eqref{eqn26} are translation invariant,
namely they satisfy $c_x(\eta)=c_0(\tau_x\eta)$.  With this choice,
the unique reversible measure $\mu_N$ is the Bernoulli measure with
parameter $p=\frac{A_1}{A_1+A_2}$, \cite{GJLV}.

We emphasize that periodic boundary conditions are crucial for the
validity of \eqref{eqn26} with a nontrivial $h$. In this special case
there are no long range correlations. In Section~\ref{s:mg+k} we show
that if the rates $c_x$ are not of type \eqref{eqn26} then -
generically - there are long range correlations.

\subsection{Totally asymmetric exclusion process}
\label{s:tasep}

The one dimensional totally asymmetric exclusion process is the
particular case of the one-dimensional asymmetric exclusion process introduced in
Section~\ref{s:2.2.5} in which particles jump only to the right. 
As discussed there, in the case of periodic boundary conditions, the
invariant measures are the Bernoulli measures with any density. We
instead consider here the boundary driven model.  
As usual we set $\Lambda_N=\{1,\cdots,N - 1\}$ and we let $\lambda_0$ and
$\lambda_1$ be the chemical potentials of the two reservoirs.  
The bulk jump rates are 
\begin{equation}
\label{jrtasep}
c_{x,x+1}(\eta) = \eta_x \, (1-\eta_{x+1}) 
\,,\qquad 
c_{x+1,x}(\eta) = 0 
\qquad\qquad x=1,\cdots, N-2
\end{equation} 
while the boundary rates are 
\begin{equation}
\label{jrtasepb}
c_{0,1}(\eta)=\eta_1 e^{\frac{\lambda_0}{2}}\,,
\;\;\qquad 
c_{N-1,N}(\eta)=\eta_Ne^{-\frac{\lambda_1}{2}}
\,,\;\;\qquad 
c_{1,0}(\eta)= c_{N,N-1}(\eta)= 0 
\end{equation} 
These rates can be obtained from our standard choice by a limiting
procedure analogous to the one described to get \eqref{br2}.

The unique invariant measure for this model has an interesting
representation due to Duchi and Schaeffer \cite{DuSch} that we
briefly recall. We duplicate the variables by introducing new random variables 
$\xi\in \{0,1\}^{\Lambda_N}$. We then define a joint distribution $\nu_N$ for
the variables $(\eta, \xi)$ as follows.
Let 
\begin{equation*}
E_x := \sum_{z=1}^x\left(\eta_z+\xi_z\right)- x 
\,,\qquad 
x=1,\cdots, N-1 
\end{equation*} 
and $E_0:= 0$.  
The measure $\nu_N$ gives positive weight only to 
\emph{complete configurations}, defined by the conditions 
\begin{equation}
\label{complete}
E_{N-1} = 0\,,\qquad \qquad 
E_x \ge 0 \,,\quad  x=1,\cdots, N-1 
\end{equation} 

Given a complete configuration we give some labels to the lattice
sites according to the following rules:
\begin{itemize}
\item[]{$x\in \Lambda_N$ has label $W$ if $\xi_x=0$ and
$E_{x-1}=E_x=0$;}
\item[]{$x\in \Lambda_N$ has label $B$ if $\xi_x=1$, $E_{x-1}=0$ and
  there are no sites on the left of $x$ labeled $W$.}
\end{itemize}

Let us denote by $N_{W}=N_{W}(\eta,\xi)$ the number of
sites with label $W$ for the complete configuration
$(\eta,\xi)$ and by $N_{B}=N_{B}(\eta,\xi)$ the number of
sites with label $B$. The measure $\nu_N$ is then defined as
\begin{equation}
\label{invtasep}
\nu_N(\eta,\xi)=\frac{1}{Z_N} \exp\Big\{ N_{W} \, \lambda_1/2 -
N_{B} \, \lambda_0/2 \Big\} 
\end{equation}
where $Z_N=Z_N(\lambda_1,\lambda_2)$ is the appropriate 
normalization constant.

The invariant measure of the boundary driven totally asymmetric exclusion
process is then the first marginal of the measure $\nu_N$, i.e.\ 
\begin{equation}
\label{margi}
\mu_N(\eta)= \sum_{\xi\in \{0,1\}^{\Lambda_N}} \nu_N (\eta ,\xi )
\end{equation} 
This result is proven by constructing a suitable 
Markov dynamics on the complete
configurations $(\eta,\xi)$ such that its projection to the $\eta$
variables coincides with the dynamics of the totally asymmetric
exclusion process.  The invariant measure of the enlarged Markov
dynamics can be easily computed and yields \eqref{invtasep}.

\section{Macroscopic theory}
\label{s:3}

As previously stated, an issue that we want to discuss is the
asymptotic behavior of the invariant measure $\mu_N$. Let us first
briefly recall the situation of reversible models. For definiteness
consider a stochastic lattice gas with reservoirs at the boundary and
assume that the chemical potential $\lambda_0$ of the boundary
reservoirs is constant and that there is no external field. As
discussed in Section~\ref{s:2.1} the unique invariant measure is the
grand-canonical Gibbs distribution
\begin{equation}\label{gibbsalb}
\mu_N^{\lambda_0}(\eta) = \frac 1 {Z_N(\lambda_0)} 
\exp\Big\{ - \mc H(\eta) + \lambda_0 
\sum_{x\in\Lambda_N}\eta_x \Big\} 
\end{equation}
where, letting $\Sigma_{N,k} := \big\{ \eta\in X^{\Lambda_N} \,\big|\:  
\sum_{x\in\Lambda_N}  \eta_x= k \big\}$, 
the grand-canonical partition function $Z_N(\lambda)$ is
\begin{equation}
Z_N(\lambda) =\sum_{k\ge 0} \, e^{\lambda \, k}  \: \sum_{\eta\in\Sigma_{N,k}}
e^{-  \mc H (\eta) }
\end{equation}

We then define $p_{0} (\lambda)$ as 
\begin{equation}
p_{0}(\lambda) := \lim_{N\to\infty} \frac 1{|\Lambda_N|} 
\log \mu_N^{\lambda_0} \Big( e^{\lambda \sum_{x\in\Lambda_N}\eta_x}
\Big) 
\end{equation}
where $|\Lambda_N|$ is the number of sites in
$\Lambda_N$.  Note that $p_0$ can be easily related to the pressure.
Let in fact $\bar p_{0}(\lambda) := \lim_{N\to\infty} |\Lambda_N|^{-1}
\log Z_N(\lambda)$ be the pressure, then $p_0(\lambda) = \bar
p_0(\lambda_0+ \lambda) - \bar p_0 (\lambda_0)$.

We then define the free energy $f_0$ as the Legendre transform of 
$p_{0}$, i.e.\ 
\begin{equation}
f_0(\rho) := \sup_{\lambda \in\bb R} \big\{ \lambda \, \rho -
p_{0}(\lambda) \big\}
\end{equation}
According to the normalization chosen $f_0$ is a convex function which
takes its minimum at the density associated to the chemical potential
$\lambda_0$, i.e.\ at $\rho_0=p_0'(0)={\bar p}'_0(\lambda_0)$.
Moreover $f_0(\rho_0)=0$.

According to the Einstein fluctuation formula \cite{E,LL}, see also
Lanford's lectures  \cite{Lan} for a complete mathematical treatment, the free
energy $f_0$ gives the asymptotic probability of observing a
fluctuation of the density, namely 
\begin{equation}
\label{level1}
\mu^{\lambda_0}_N \Big( \frac 1{|\Lambda_N|} \sum_{x\in\Lambda_N} \eta_x
\approx \rho \Big) \sim \exp\big\{ - |\Lambda_N| \, f_0(\rho) \big\}
\end{equation}
here $a \approx b$ means  closeness in $\bb R$ and
$\sim$ denotes logarithmic equivalence as $|\Lambda_N|$ diverges. 

In discussing nonequilibrium models, which are not translationally
invariant, it is important to establish a generalization of the above
fluctuation formula. We want to compute the asymptotic probability of
a fluctuation not of the average density but of the density profile.  
In fact, already Einstein \cite{E}, considered density profiles in
small fluctuations from equilibrium. 
We introduce the empirical density as follows.  To each
microscopic configuration $\eta\in X^{\Lambda_N}$ we associate a
macroscopic profile $\pi^N(u)=\pi^N(\eta;u)$, $u\in \Lambda$, by
requiring that for each smooth function $G:\Lambda \to \bb R$
\begin{equation}
\label{emden}
\langle \pi^N, G \rangle = 
\int_{\Lambda} \!du\: \pi^N(u) \, G(u) 
= \frac{1}{N^{d}} \sum_{x\in\Lambda_N} G(x/N) \eta_x 
\end{equation}
so that $\pi^N(u)$ is the local density at the macroscopic point
$u=x/N$ in $\Lambda$. Let $\rho=\rho(u)$ be a given density profile. 
Then \eqref{level1} can be recast as 
\begin{equation}
\label{level2}
\mu^{\lambda_0}_N \big( \pi^N \approx \rho  \big) 
\sim \exp\big\{ - N^d \, \mc F_0(\rho) \big\}
\end{equation}
Here $ \rho \approx \rho' $ means that their averages over macroscopically small neighborhoods 
are close and $\mc F_0(\rho)$ is the local and convex functional 
\begin{equation}
\mc F_0(\rho) = \int_{\Lambda} \!du \: f_0\big(\rho(u)\big) 
\end{equation}

\medskip
For non-reversible systems we shall look for a fluctuation formula
like \eqref{level2} which, in the same spirit as Einstein, we shall
consider as the definition of the nonequilibrium free energy.  While
in the reversible setting discussed above the invariant measure
$\mu_N$ is given by the Gibbs distribution \eqref{gibbsalb}, in a
non-reversible system $\mu_N$ is not, in general, explicitly
known. For special models, powerful combinatorial methods have been
used \cite{DLS1,DLS2,DLS3,DE}.  In the sequel we shall discuss instead
the strategy introduced in \cite{BDGJL1,BDGJL2} which is based on the
following idea.  As $N$ diverges the evolution of the thermodynamic
variables is described by a closed macroscopic evolution called
hydrodynamic equation.  The microscopic details are then encoded in
the transport coefficients appearing in the hydrodynamic equation.  In
the cases discussed here, these transport coefficients are the
diffusion coefficient and the mobility.  For the Glauber + Kawasaki
dynamics also the reaction rates are involved.  We then compute the
asymptotic probability of fluctuations from the typical hydrodynamical
behavior generalizing to a dynamical setting the Einstein fluctuation
formula \eqref{level2}.  The nonequilibrium free energy $\mc F$ is
then characterized as the solution of a variational problem, from
which we derive a Hamilton-Jacobi equation involving the transport
coefficients.  This is an infinite-dimensional strategy analogous to
the Freidlin--Wentzell theory for diffusion processes, \cite{FV}.

Of course, in the case of reversible systems, the solution to the
Hamilton-Jacobi equation coincides with the equilibrium free energy
$\mc F_0$.  This is essentially the characterization of $\mc F_0$
given by Onsager-Machlup \cite{OMA}, extended to a non-linear context.

In Section~\ref{s:was} we discuss the hydrodynamics and the associated
dynamical large deviations principle of weakly asymmetric lattice
gases.  In Section~\ref{s:hj} we recall the derivation of the
Hamilton-Jacobi equation and we discuss the form of the nonequilibrium
free energy for the specific models introduced in Section~\ref{s:2}.
We will also discuss a toy model for the invariant measure of the KMP
process.  In Section~\ref{s:mgf} we obtain macroscopic equation
satisfied by the correlation functions and we discuss whether
correlations are positive or negative.  In Section~\ref{s:mwas} we
show that for weakly asymmetric lattice gases with periodic boundary
conditions the nonequilibrium free energy coincides with the
equilibrium one.  In Section~\ref{s:mg+k} we discuss the macroscopic
property of the Glauber + Kawasaki dynamics \cite{BJ,GJLV}.  Finally,
in Section~\ref{s:mtasep}, starting from the results in \cite{DuSch},
we show how the representation for the nonequilibrium free energy of
the totally asymmetric exclusion process obtained in \cite{DLS3} can
be formulated as a minimization problem.

\subsection{Hydrodynamics and dynamical large deviations}
\label{s:was}

We consider an asymmetric model as defined by the rates
\eqref{ratas}. If the microscopic external field $F$ is of order $1$,
the appropriate scaling is the Euler one, i.e.\ both space and time
are rescaled by a factor $N$, and the hydrodynamic equation is given
by an hyperbolic equation, see \cite{KL} and references therein.  
We here consider instead the case in which the external field is of
the order $1/N$ as in \eqref{wasym}.  Then the hydrodynamic limit is
obtained in the diffusive scaling and given by a parabolic equation.
Let $\pi^N(t)$ be the empirical density, as defined in \eqref{emden},
corresponding to the particles configuration at time $N^2t$,
$\pi^N(t,u)$ is then a random space-time trajectory; as $N\to \infty$
it converges however to a deterministic function.
Referring to \cite{KL,S,VJ} for periodic boundary conditions and to 
\cite{BDGJL2,BDGJL6,els1,els2} for open systems, we here state
the law of large numbers, as $N\to\infty$, of the empirical density
$\pi^N$ for weakly asymmetric lattice gases.  The macroscopic
evolution of the density is described by a (in general nonlinear)
diffusion equation with a transport term corresponding to the external
field, namely
\begin{equation}
\label{hyde}
\partial_t \rho = 
\nabla \cdot \Big[ \frac 12 \,D(\rho) \nabla \rho   -\chi(\rho) E \Big]
\end{equation}
where $D$ is the diffusion matrix, obtained from the microscopic
dynamics by a Green-Kubo formula \cite[II.2.2]{S}, and $\chi$ is the
mobility matrix, obtained by linear response theory \cite[II.2.5]{S}. 
In \eqref{hyde} $\cdot$ denotes the standard inner product in $\bb R^d$.
This equation has to be supplemented by the boundary conditions which are
either periodic when $\Lambda$ is the torus or the non-homogeneous Dirichlet
condition
\begin{equation}
\label{hydebc}
\lambda\big(\rho(t,u)\big) = \lambda_0(u) \;, \quad u\in\partial\Lambda 
\end{equation}
in the case of boundary driven systems. Here $\partial\Lambda$ is the
boundary of $\Lambda$, $\lambda(\rho)=f_0'(\rho)$ is the chemical
potential associated to the microscopic Hamiltonian $\mc H$, and
$\lambda_0$ is the chemical potential of the boundary reservoirs.
Finally the initial condition for \eqref{hyde} is obtained as the
limiting empirical density of the chosen microscopic initial
configuration of particles.

We obtain an equilibrium model either if $\Lambda$ is the torus and
there is no external field or in the case of boundary driven systems
in which the external field in the bulk matches the driving from the
boundary; in particular if $\lambda_0$ is constant and $E$
vanishes. In the other cases the stationary state supports a non
vanishing current and the systems is out of equilibrium.

The coefficients $D$ and $\chi$ are related by the Einstein
relation $D=R^{-1}\chi$, where $R$ is the compressibility: $R^{-1} =
f_0''$, in which $f_0$ is the equilibrium free energy associated to
the Hamiltonian $\mc H$, see \cite{S}.  
For gradient lattice gases, as defined in Section~\ref{s:2.2}, the
diffusion matrix $D$ and the mobility $\chi$ are multiples of the
identity. For non gradient models in general $D$ and $\chi$ are not diagonal,
however, as shown in \cite[Lemma~8.3]{VJ}, if the Hamiltonian $\mc H$ is
invariant w.r.t.\ rotation of $\pi/2$, then $D$ and $\chi$ are
diagonal.  

\bigskip
We next discuss the large deviation properties of the empirical
density; the derivation can be found in \cite{BDGJL2,BDGJL4,KL,S}. 
Fix a smooth trajectory $\hat\rho\equiv \hat\rho(t,u)$,
$(t,u)\in[0,T]\times\Lambda$. We want to compute the asymptotic
probability that the empirical density $\pi^N$  is in a 
small neighborhood of $\hat\rho$. If $\hat\rho$ is not a solution to
\eqref{hyde}, this probability will be exponentially small and the
corresponding rate is called the \emph{large deviation dynamical rate
  functional}.

Consider an initial configuration $\eta$ whose empirical measure
approximates, as $N$ diverges, $\hat\rho(0)$ and let $\bb P^N_\eta$ be
the law of the microscopic process starting from such initial
condition.  The dynamical large deviation principle for the empirical
density states that
\begin{equation}
\label{f1}
\bb P^N_\eta \big( \mc \pi^N \approx \hat\rho \big) 
\sim \exp\big\{ - N^d \, I_{[0,T]}(\hat\rho) \big\}
\end{equation}
where the rate functional $I_{[0,T]}$ is 
\begin{equation}
\label{Ic}
I_{[0,T]}(\hat\rho)\;=\; \frac 12 \int_0^T \!dt \,
\big\langle  \nabla H , \chi(\hat\rho) \nabla H \rangle
\end{equation}
in which $\langle\,,\,\rangle$ denotes integration in the space
variables and $\nabla H\equiv \nabla H(t,u)$ is
the extra gradient external field needed to produce the
fluctuation $\hat\rho$, namely such that
\begin{equation}
\label{hydef}
\partial_t \hat\rho =  
\nabla \cdot \Big[ \frac 12 D(\hat\rho) \nabla \hat\rho   
-\chi(\hat\rho) (E+\nabla H) \Big]
\end{equation}

The interpretation of \eqref{Ic} is straightforward; since $\chi$ is
the mobility, $I_{[0,T]}(\hat\rho)$ is the work done by the external
field $\nabla H$ to produce the fluctuation $\hat\rho$ in the time
interval $[0,T]$.

\subsection{Thermodynamic functionals and Hamilton-Jacobi equation}
\label{s:hj}

Consider the following physical situation.
The system is macroscopically in the stationary profile 
$\bar\rho\equiv\bar\rho(u)$, $u\in\Lambda$ 
(a stationary solution to \eqref{hyde}) at $t=-\infty$, but at $t=0$ 
we find it in the state $\rho$.  
We want to determine the most probable trajectory followed in the
spontaneous creation of this fluctuation. According to \eqref{f1} this
trajectory is the one that minimizes $I$ among all trajectories
$\hat\rho(t)$ connecting $\bar\rho$ to ${\rho}$ in the time interval
$[-\infty,0]$. We thus define the so-called \emph{quasi-potential} as 
\begin{equation}
\label{quasipot}
V(\rho)  = \inf_{\substack{\hat\rho \,:\: \hat\rho(-\infty)=\bar\rho\\
 \hat\rho(0)=\rho }} I_{[-\infty,0]} (\hat\rho)
\end{equation}
As shown in \cite{BDGJL2,BDGJL3}, the functional $V$ solves the
Hamilton-Jacobi equation
\begin{equation}
\label{HJ}
\frac 12 \Big \langle \nabla \frac {\delta V}{\delta \rho}  , 
\chi (\rho) \nabla \frac {\delta V}{\delta \rho} \Big \rangle
+ \Big \langle \frac {\delta V}{\delta \rho} , 
\nabla \cdot  
\Big[  \frac 12 D(\rho) \nabla \rho  - \chi(\rho) E \Big] \Big\rangle 
= 0
\end{equation}
note that there is no uniqueness of solutions, e.g.\ $V=0$ is always a
solution. In \cite{BDGJL2} it is discussed the appropriate
selection criterion, that is $V$ is the maximal solution to \eqref{HJ}.

If the system is in equilibrium then the quasi-potential $V$ coincides
with the variation of the equilibrium free energy associated to the
profile $\rho$.  The latter can be characterized, by the Einstein
fluctuation formula, as the rate of the asymptotic probability of
observing a given density profile in the equilibrium measure. Namely,
if $\mu_N$ is the invariant measure of the generator $L_N$, then
\begin{equation}
\label{ldeq}
\mu_N(\pi^N \approx \rho \big) \sim \exp\big\{ - N^d \, V(\rho) \big\}
\end{equation}
This relation holds also for nonequilibrium systems, see
\cite{BDGJL2,bg} and, in this sense, the solution to the variational
problem \eqref{quasipot} is the appropriate generalization of the 
free energy for nonequilibrium systems.  Finally, as
discussed in \cite{BDGJL2}, for \emph{generic} nonequilibrium models
the quasi-potential $V$ is a \emph{non-local} functional of
$\rho$. Notable exceptions are the zero-range model and the case,
discussed in the Section~\ref{s:mwas}, of systems with weak external
field and periodic boundary conditions.  We next recall some results
on the quasi potential for specific lattice gases.

\bigskip
\noindent\emph{Zero range process.} \ 
We consider the zero range process as introduced in
Section~\ref{s:2.3} either in the torus or in a bounded domain with a
weak external field $E$. Recalling \eqref{INV} and \eqref{Z=}, 
we define the function $\Phi:\bb R_+\to\bb R_+$ as 
the activity corresponding to the density $\alpha$, i.e.\ such that
\begin{equation}
\label{f-1}
\alpha = \frac{1}{Z(\Phi(\alpha))} \:
 \sum_{k=0}^\infty  \:k  \; \frac{\Phi(\alpha)^k}{g(k)!} 
\end{equation}
where $Z(\varphi)$ is defined in \eqref{Z=}.
In other words $\alpha\mapsto \Phi(\alpha)$ 
is the inverse of the function $\varphi \mapsto R(\varphi)$ defined by
\begin{equation}
\label{Rf}
R(\varphi) = \varphi \, \frac {Z'(\varphi)}{Z(\varphi)}
\end{equation}
As shown in \cite{BDGJL1,BDGJL2,DFzr,KL},
the hydrodynamic equation for the zero range process is then
\eqref{hyde} with $D=\Phi'$ and $\chi=\Phi$. In the case of
independent random walks, i.e.\ $g(k)=k$, $\Phi$ is the identity so that $D=1$ and
$\chi(\rho)=\rho$. 

Since for the zero range process, as discussed in Section~\ref{s:2.3}, the
invariant measure is always a product measure, the quasi potential $V$
is a local functional. Its form can be computed directly from the
invariant measure by requiring that \eqref{ldeq} holds. On the other hand
it is also possible to solve explicitly the Hamilton-Jacobi equation
\eqref{HJ}. As shown in \cite{BDGJL1,BDGJL2} we get
\begin{equation}
V(\rho)=
\int_\Lambda \!du \: 
\left[ \rho(u) \log \frac {\Phi(\rho(u))}{\bar\varphi(u)}  
- \log \frac{Z(\Phi(\rho(u)))}{Z(\bar\varphi(u))}
\right]
\label{E}
\end{equation}
where $\bar\varphi(u)=\Phi(\bar\rho(u))$ is the stationary activity profile, 
$\bar\rho$ being the stationary solution to \eqref{hyde}, i.e.\ the
stationary density profile. Equivalently $\bar\varphi$ solves 
\begin{equation}
\label{acst}
\left\{
\begin{array}{ll}
\frac 12 \Delta \bar\varphi - \nabla \cdot \big( \bar\varphi E \big) 
= 0 & u \in \Lambda 
\\ 
\bar\varphi(u) = \exp\{ \lambda_0(u) \}   
& u \in \partial \Lambda 
\end{array}
\right.
\end{equation}
which, recalling \eqref{wasym}, is just the continuos 
limit of \eqref{laminv}.

\bigskip
\noindent\emph{Boundary driven symmetric exclusion process.} \ 
We consider here the one-dimensional symmetric exclusion process as
introduced in Section~\ref{s:2.2.5} with
$\Lambda = (0,1)$. Let $\rho_0$ and $\rho_1$ be the boundary
densities. As shown in \cite{BDGJL2,els1,els2}, 
the hydrodynamic equation is \eqref{hyde} with $D=1$ and
$\chi(\rho)=\rho (1-\rho)$. For this model, if $\rho_0\neq\rho_1$ 
the quasi potential is non local, which is the signature of macroscopic long
range correlations.
The quasi potential cannot be written in a closed form, but can be
obtained by solving a one-dimensional boundary value problem. This has
been proven in \cite{DLS1,DLS2} by combinatorial methods and in 
\cite{BDGJL2,BDGJL4} by the dynamical/variational approach here
presented. The result is the following. 
\begin{equation}
\label{qpes}
V(\rho) =\sup_{f} \int_0^1\!du \: 
\Big[ \rho \log \frac{\rho}{f}  + (1-\rho) \log \frac{1-\rho}{1-f}  
+ \log \frac{f'}{\rho_1-\rho_0} \Big] 
\end{equation}
where the supremum is carried out over all strictly monotone smooth functions
$f$ satisfying the boundary conditions $f(0)=\rho_0$, $f(1)=\rho_1$. 
It has also been shown that there exists a unique maximizer for the
variational problem \eqref{qpes} which is the unique strictly 
monotone solution to the non-linear boundary value problem 
\begin{equation}
\label{deq}
\left\{
\begin{array}{l}
f (1-f) \frac{f''}{(f')^2} +f = \rho   \\
f(0)=\rho_0\,,\; f(1)=\rho_1
\end{array}
\right.
\end{equation}
in which $\rho=\rho(u)$ is the prescribed fluctuation. 
Knowing that \eqref{qpes} is the answer, the proof amounts to some
lengthy but straightforward computations in showing that it solves the
Hamilton-Jacobi equation \eqref{HJ}, see \cite{BDGJL4} for the
details. From \eqref{qpes}, since $V$ is
expressed as the supremum of convex functionals we get ``for free''
that $V$ is a convex functional. However, as shown below, this
convexity property does not hold in general.

Variational formulae like \eqref{qpes} are typical in statistical
mechanics, but here the interpretation it is rather unclear. Firstly
it appears strange that we need to maximize and not to minimize, secondly
the meaning of the test function $f$ is not apparent. 
For the second issue we mention that a dynamical interpretation of $f$
in terms of the hydrodynamics of the time reversed process is discussed
in \cite{BDGJL2}. For the first issue we shall show that it is
connected with the convexity properties of the mobility $\chi$.

We mention that an expression similar to \eqref{qpes} has also been
obtained for the boundary driven weakly asymmetric exclusion process
in \cite{DE}. Also the Hamilton-Jacobi approach can be applied
successfully, see \cite{BGL}.

\bigskip
\noindent\emph{Boundary driven KMP process.} \ 
We consider here the KMP process introduced in Section~\ref{s:kmp}.
The hydrodynamic equation is \eqref{hyde} with $D=1$ and
$\chi(\rho)=\rho^2$. Note that here $\rho$ is the energy density and
not the particle density as for lattice gases. 
Similarly to the boundary driven symmetric exclusion process, as shown in
\cite{bgl}, the quasi potential can be obtained by solving a
one-dimensional boundary value problem.
The result is the following. 
\begin{equation}
\label{qpkmp}
V(\rho) =\inf_{f} \mc G(\rho,f)
\end{equation}
where
\begin{equation}
\label{toymrf}
\mc G (\rho,f) = \int_0^1\!du \: 
\Big[ \frac{\rho}{f} -1 -\log \frac{\rho}{f} 
- \log \frac{f'}{T_1-T_0} \Big] 
\end{equation}
and the infimum is carried out over all strictly monotone smooth functions
$f$ satisfying the boundary conditions $f(0)=T_0$, $f(1)=T_1$. 
It has also been shown that there exists a unique minimizer for the
variational problem \eqref{qpkmp} which is the unique strictly 
monotone solution to the non-linear boundary value problem 
\begin{equation}
\label{deqkmp}
\left\{
\begin{array}{l}
f^2 \frac{f''}{(f')^2} -f = -\rho   \\
f(0)=T_0\,,\; f(1)=T_1
\end{array}
\right.
\end{equation}
in which $\rho=\rho(u)$ is the prescribed fluctuation. 
As the for the boundary driven symmetric exclusion process, knowing 
that \eqref{qpkmp} is the answer, the proof amounts to some
lengthy but straightforward computations in showing that it solves the
Hamilton-Jacobi equation \eqref{HJ}. 
Unlike the boundary driven symmetric exclusion process, the quasi
potential for the KMP process is not convex.

A possible interpretation of \eqref{qpkmp} is the following.  The
local functional $\mc G(\rho,f)$ can be thought of as a joint rate
functional for both the energy density $\rho$ and the function $f$,
which we can interpreted as a temperature profile.  Then the
minimization procedure of \eqref{qpkmp} corresponds to the application
of a contraction principle.  We therefore search for the best hidden
temperature profile $f$ associated to the energy density profile
$\rho$.  This is the inspiring idea behind the following toy model for
the invariant measure.

\bigskip
We will show that the functional $V$ in \eqref{qpkmp} is the large
deviations rate functional of a measure on $\bb R_+^{\Lambda_N}$ which
is ``simple'' enough to be described explicitly and ``rich'' enough to
produce such a non-local rate functional. Recall that in Section~\ref{s:kmp} we
have obtained an explicit representation of the invariant measure of
the KMP process with a single oscillator as a convex combinations of
exponential distributions. 

We assume $T_0\le T_1$ and  
let $t_1,\cdots, t_{N-1}$ be independent uniform random variables 
on the interval $[T_0,T_1]$. Denote by 
$t_{[1]}\le  t_{[2]}\le \cdots \le  t_{[N-1]}$ be order statistics of
$t_1,\cdots, t_{N-1}$, i.e.\ $t_{[1]}$ is the smallest among the
$t_i$, $t_{[2]}$ the second smallest and so on. Denote by $\varrho_N$,
the distribution of the random vector $t_{[1]},\cdots,t_{[N-1]}$; note that 
$\varrho_N$ is a probability on $[T_0,T_1]^{N-1}$. We then
define $\nu_N$ as the probability measure on $\bb R_+^{\Lambda_N}$
whose density w.r.t.\ the Lebesgue measure 
$d\xi =\prod_{x\in\Lambda_N} d\xi_x$ is given by 
\begin{equation}
\label{toym}
\frac{d \nu_N}{d\xi}  
= \int\! \varrho_N(dt_1,\cdots,dt_{N-1}) 
\: \prod_{x\in\Lambda_N} \frac{1}{t_x} \exp\{ - \xi_x/t_x\}
\end{equation}
That is $\nu_N$ is a mixture of the exponential Gibbs distribution 
with temperature profile $T(x/N)=t_x$. The measure $\nu_N$ is not the
invariant measure of the KMP process; if we compare \eqref{toym} for a
single oscillator, $N=2$, with the exact expression in \eqref{4.2} we
see that we replaced the arcsin distribution in $[T_0,T_1]$ with
the uniform one. As $N$ diverges, the measure $\nu_N$ is however a
good approximation of the true invariant measure in the sense that it
leads the rate function in \eqref{qpkmp}. In particular it has the
correct asymptotic form of the two point correlations. 

To prove the above statement, let us consider the probability measure 
$\tilde{\nu}_N$ on the space $\bb R_+^{\Lambda_N}\times
[T_0,T_1]^{\Lambda_N}$ given by  
\begin{equation}
\label{toym2}
\tilde{\nu}_N(d\xi,dt) =  \varrho_N(dt) 
\: \prod_{x\in\Lambda_N} \frac{1}{t_x} 
\exp\{ - \xi_x/t_x\} \,d\xi_x
\end{equation}
so that $\nu_N$ in \eqref{toym} is obtained as the first marginal of
$\tilde{\nu}_N$, i.e.\ integrating on the second variable $t$. 
Recalling the definition \eqref{emden} of the empirical density
$\pi^N$, we likewise define the empirical temperature profile $\tau^N$
by requiring that for each smooth function $G$ on $\Lambda$
$\langle \tau^N, G \rangle = \frac 1N \sum_{x\in\Lambda_N} G(x/N) t_x $
Given a smooth function $\rho:\Lambda\to \bb R_+$ and a smooth strictly 
increasing function $f: \Lambda\to [T_0,T_1]$ such that 
$f(0)=T_0$ and $f(1)=T_1$, we claim that 
\begin{equation}
\label{toymld}
\tilde{\nu}_N\big(\pi^N \approx \rho\,,\, \tau^N\approx  f \big) 
\sim \exp\big\{ - N \mc G (\rho,f) \big\}
\end{equation}
where $\mc G$ was defined in \eqref{toymrf}.
To obtain this result, we first observe that if $e_1,\cdots,e_N$ are
$N$ independent exponential random variables with parameter $T$, then 
\begin{equation*}
\bb P \Big( \frac 1N \sum_{i=1}^N e_i \approx \alpha \Big) \sim 
\exp\big\{ - N\, \big[\alpha/T -1 - \log (\alpha/T) \big] \big\}
\end{equation*}
We also recall, see e.g.\ \cite[I.6]{Feller}, that the random
variables $\Delta_1 := t_{[1]}-T_0,\,\Delta_2 := t_{[2]}-t_{[1]},
\cdots,\Delta_N := T_1-t_{[N-1]}$ are distributed according the
product of $N$ exponential \emph{conditioned} on
$\Delta_1+\cdots+\Delta_N=T_1-T_0$.  We then get
\begin{equation*}
\varrho_N \big( \tau^N \approx f \big) \sim 
\exp\Big\{ -N 
 \int_0^1 \!du \: \Big[ - \log \frac {f'(u)}{T_1-T_0} \Big]\Big\}
\end{equation*}
Since, conditionally on the random variables $t_{[x]}$, $x\in\Lambda_N$, 
the distribution of $\xi$ is the product of exponentials, \eqref{toymld}
follows.
Finally, from \eqref{toymld}, by maximizing over the possible 
values of $f$, we easily get that 
\begin{equation}
\label{toymld1}
\nu_N\big(\pi^N \approx \rho\big) 
\sim \exp\big\{ - N \inf_f \mc G (\rho,f) \big\}
\end{equation}

\subsection{Macroscopic correlation functions}
\label{s:mgf}

In Sections~\ref{s:2.2.5} and \ref{s:kmp} we found exact formulae for
the two-point correlation functions $C_N(x,y)=\mu_N(\eta_x;\eta_y)$ of
the invariant measure $\mu_N$, both for the one dimensional boundary
driven symmetric exclusion process, see \eqref{covSEP}, 
and for the boundary driven KMP process, see \eqref{covKMP}.
For both models we found that, out of equilibrium, they admit long
range correlations of order $1/N$.  More precisely we have,
$$
C_N(x,y)=\frac 1N C\Big(\frac xN,\frac yN\Big) + O\Big(\frac 1{N^2}\Big)
$$
with
\begin{equation}\label{alb1}
C(u,v) = C(v,u) =  -(\rho_1-\rho_0)^2 u(1-v)\,\,,\,\,\,\,0\leq u<v\leq 1
\end{equation}
for the boundary driven symmetric exclusion process, and
\begin{equation}\label{alb2}
C(u,v) = C(v,u) = +(T_1-T_0)^2 u(1-v)\,\,,\,\,\,\,0\leq u<v\leq 1
\end{equation}
for the boundary driven KMP process.
Notice that the above functions \eqref{alb1} and \eqref{alb2}
only differ by a sign. Moreover the off-diagonal part of the
above covariance is proportional to the Green function of the
Laplacian on the interval $[0,1]$ with Dirichlet boundary conditions,
\begin{equation}\label{green}
\Delta^{-1}(u,v)=\Delta^{-1}(v,u)= -  u(1-v)\,\,,\,\,\,\,0\leq u\leq v\leq1
\end{equation}
namely the solution to the problem
$\partial_u^2\, \Delta^{-1} (u,v) = \delta(u-v),\,0\leq u,v\leq1$,
with boundary condition $\Delta^{-1}(u,v)=0$ 
if either $u$ or $v$ is $0$ or $1$.

In this section we will derive the above results from a purely
macroscopic point of view.  More precisely, we consider a
one-dimensional boundary driven system with $\Lambda=(0,1)$ and no
external field and we assume that the transport coefficients in
\eqref{hyde} are of the following form. The diffusion coefficient is
constant, we set $D(\rho)=1$, and $\chi(\rho)$ is quadratic so that
$\chi''$ is constant.  We show that such models have positive,
resp. negative, correlations if $\chi''\ge0$, resp. $\chi''\le0$.

Recall that the quasi-potential $V(\rho)$ solves the Hamilton-Jacobi
equation \eqref{HJ}, which in this context reads
\begin{equation}\label{HJnoE}
\Big\langle \nabla \frac {\delta V}{\delta \rho}  , 
\chi (\rho) \nabla \frac {\delta V}{\delta \rho} - \nabla \rho \Big\rangle = 0
\end{equation}
The functional $V$ assumes its minimum at $\bar\rho$, the stationary
solution to \eqref{hyde}, which in this case is a linear function.
The correlation function $C(u,v)$, which measures the covariance of
the density fluctuations with respect to the invariant measure, is
then obtained in the quadratic approximation of $V$, i.e.\
\begin{equation}\label{expV}
V(\rho) = \frac12 \big\langle
(\rho-\bar\rho),C^{-1}(\rho-\bar\rho)\big\rangle +
O\big((\rho-\bar\rho)^3\big)
\end{equation}
where $C^{-1}$ denotes the inverse operator of $C$.
We can therefore get an equation for $C(u,v)$ by expanding the
Hamilton-Jacobi equation \eqref{HJnoE} up to second order in
$(\rho-\bar\rho)$.

It is convenient to introduce the
``pressure" $G(h)$, see \cite{BDGJL2}, defined as the Legendre
transform of the quasi-potential $V(\rho)$,
$$
G(h) = \sup_\rho \Big\{\langle h,\rho\rangle - V(\rho) \Big\}
$$
Here $h=h(u)$ can be interpreted as a chemical potential profile.
By Legendre duality, equation \eqref{HJnoE} can be rewritten as the following
Hamilton-Jacobi equation for the pressure,
\begin{equation}\label{HJG}
\Big\langle \nabla h , 
\chi \Big(\frac {\delta G}{\delta h}\Big) \nabla h 
- \nabla \frac {\delta G}{\delta h} \Big\rangle = 0
\end{equation}
for any $h$ which satisfies the boundary conditions $h(0)=h(1)=0$.
Moreover the expansion \eqref{expV} gets translated into the following
expansion for $G$,
\begin{equation}\label{expG}
G(h) = \langle h,\bar\rho \rangle + \frac12 \langle h,C h\rangle + O(h^3)
\end{equation}
Hence the macroscopic correlation function $C$ can be obtained by expanding
equation \eqref{HJG} up to second order in $h$.

From equation \eqref{expG} we have
\begin{equation}\label{expG1}
\frac{\delta G}{\delta h(u)} = \bar\rho(u) + (C h)(u) + O(h^2)
\end{equation}
If we thus plug \eqref{expG1} into \eqref{HJG} and neglect the terms of
order $h^3$ we get
\begin{equation}\label{alb3}
\Big\langle \nabla h ,  \chi (\bar\rho) \nabla h 
-\nabla Ch \Big\rangle = 0
\end{equation}
for all chemical potential profiles $h$ such that $h(0)=h(1)=0$.  To
derive the above equation we used the fact that $\bar\rho$ is linear.
The macroscopic correlation function $C$ can then be determined as the
solution to equation \eqref{alb3} satisfying the boundary condition
$C(u,v)=0$ if $u\neq v$ and either $u$ or $v$ is $0$ or $1$.  This
condition is due to the fact that the values of the density at the
boundary is fixed by the reservoirs.

We next define the nonequilibrium contribution to the covariance as
the function $B$ such that
\begin{equation}\label{alb4}
C(u,v) = \chi(\bar\rho(u))\delta(u-v) + B(u,v)\,\,,\,\,\,\, u,v\in\Lambda
\end{equation}
Note that, since $D=1$, $\chi(\bar\rho(u))$ is the local equilibrium
variance.  
By plugging \eqref{alb4} into \eqref{alb3}, we get that $B$ solves
\begin{equation}\label{alb7}
(\partial_u^2+\partial_v^2)B(u,v) = - (\nabla\bar\rho)^2\,
  \chi^{\prime\prime}\, \delta(u-v) 
\end{equation}
together with the boundary condition $B(u,v)=0$ if either $u$ or $v$
is $0$ or $1$.  The above equation can also be derived within the
fluctuating hydrodynamic theory, see \cite{S1}.
Hence
$$
B(u,v) =  - \frac 12 (\nabla\bar\rho)^2\,\chi^{\prime\prime}\,\Delta^{-1}(u,v)
$$
which, by \eqref{green} and recalling that $\chi(\rho)=\rho(1-\rho)$
for the exclusion process and $\chi(\rho)=\rho^2$ for the KMP process,
agrees with \eqref{alb1} and \eqref{alb2}.

\smallskip
In \cite{BDGJL10} we derive the equation satisfied by the off diagonal
covariance $B$ for arbitrary dimension, $D$, $\chi$, and external
field $E$. This equation allows to establish, for a class of models,
whether the correlations are positive or negative.

\subsection{Weakly asymmetric models with periodic boundary conditions} 
\label{s:mwas}

We consider here a lattice gas with periodic boundary conditions,
namely $\Lambda$ is the $d$-dimensional torus, and constant weak
external field $E$.  As discussed in Section~\ref{s:2.2}, from a
microscopic point of view, if the model is \emph{gradient} then the
invariant measure does not depend on the external field $E$. As we
show here, from a macroscopic point of view, \emph{any}
system behaves as gradient models.

The precise statement is the following. Consider the variational
problem \eqref{quasipot} defining the quasi-potential $V$ in the
present setting of periodic boundary conditions and constant external
field $E$. Then $V$ does not depend on $E$ and therefore coincides
with the solution to \eqref{quasipot} with $E=0$, namely with the
free energy associated to the microscopic Hamiltonian $\mc H$.

We suppose given the transport coefficients $D$ and $\chi$ in
\eqref{Ic}-\eqref{hydef} so that the Einstein relationship 
$D(\rho) = R(\rho)^{-1} \chi(\rho)$ holds; recall that while $D$ and
$\chi$ are matrices, the compressibility $R$ is a scalar. 
In the case of periodic boundary conditions and constant field $E$
there is a one parameter family of stationary solutions to
\eqref{hyde} which are simply the constant functions $\bar\rho(u)=m$,
$m\in\bb R_+$.  Given $m\in\bb R_+$ we define
\begin{equation*}
f_m(\rho) = \int_{m}^\rho\!dr \int_{m}^r\!dr' \frac 1{R(r')} 
\end{equation*}
which is a strictly convex function with minimum at $\rho=m$.  
We claim that the solution of the variational problem \eqref{quasipot}
with $\bar\rho=m$ is the functional
\begin{equation}
\label{floc}
\mc F_m(\rho) = \int_\Lambda\!du \: f_m(\rho(u))
\end{equation}
for any value  of the external field $E$. 

If $E=0$, by using the Einstein relation
$D(\rho)=f_m^{\prime\prime}(\rho)\chi(\rho)$, it is easy to check that
$\mc F_m$ solves the Hamilton-Jacobi \eqref{HJ}. If $E$ is a constant,
since the boundary conditions are periodic, we have that
\begin{equation*}
\big \langle \frac {\delta \mc F_m}{\delta \rho} , 
\nabla \cdot  \chi(\rho) E \big\rangle =0 
\end{equation*}
hence $\mc F_m$ solves the Hamilton-Jacobi \eqref{HJ} for any (constant) 
external field $E$. 
It is also not difficult to check that $\mc F_m$ is the maximal
solution to the Hamilton-Jacobi equation \eqref{HJ}
so that the claim is proven.

\subsection{Glauber + Kawasaki}
\label{s:mg+k}

We consider here the macroscopic behavior of the Glauber + Kawasaki
process introduced in Section~\ref{s:2.4}. The empirical density is
defined as in \eqref{emden}. We emphasize that in this model the
empirical density is not locally conserved due the reaction terms in
the microscopic dynamic \eqref{eqn1}. Accordingly, the hydrodynamic
equation is given by the reaction diffusion equation 
\begin{equation}
\label{eqn6}
\partial_t \rho = \frac{1}{2}\Delta \rho + b(\rho)-d(\rho) 
\end{equation}
where the reaction
 terms $b$ and $d$, which are polynomials in $\rho$, 
are determined by the rates $c_x(\eta)$ in \eqref{eqn1} as follows, \cite{DFL,JLV},
\begin{equation}\label{giada1}
b(\rho) ={\nu_\rho}\big(c_0(\eta)(1-\eta_0)\big)\,\,,\,\,\,\,
d(\rho) = {\nu_\rho}\big(c_0(\eta)\eta_0\big)
\end{equation}
where $\nu_\rho$ is the Bernoulli measure with density $\rho$.
In particular, in the reversible case where the rates $c_x(\eta)$ are
as in \eqref{eqn26}, $b(\rho)$ and $d(\rho)$ have the form
\begin{equation}\label{giada2}
b(\rho) = A_1 \, (1-\rho) \, \varphi(\rho)\,\,,\,\,\,\,
d(\rho) = A_2 \, \rho \, \varphi(\rho)
\end{equation}
where $\varphi(\rho)$ is the expected value of $h(\eta)$ in
\eqref{eqn26} with respect to $\nu_\rho$ and $A_1,\,A_2\ge0$.  
We consider this system only with periodic boundary conditions.  The
equilibrium profile thus corresponds to a constant density $\bar\rho$
which solves $b(\rho)=d(\rho)$ and gives an absolute minimum of the
potential $U$, defined by $U'(\rho) = -[b(\rho) - d(\rho)]$.

The associated large deviation asymptotics is in the same form as in
\eqref{f1}, but here the rate functional $I_{[0,T]}$ is not simply
quadratic in the external field. Indeed in \cite{JLV} it is proven
that it is given by 
\begin{eqnarray}
\label{eqn14}
\nonumber
I_{[0,T]}(\hat\rho) &=& 
\int_0^{T}  \!dt \, \Big\{
\frac{1}{2} \big\langle \nabla H, \hat\rho (1-\hat\rho) \nabla H \big\rangle
\\
&&
+ \Big\langle  b(\hat\rho) , \big(1-e^{H}+ H e^{H}\big) \Big\rangle 
+ \Big\langle d(\hat\rho) , \big(1- e^{-H}-H e^{-H} \big) \Big\rangle 
\Big\}
\end{eqnarray}
where the external potential $H$ is connected to the fluctuation
$\hat\rho$ by 
\begin{equation} 
\label{eqn10}
\partial_t \hat\rho = \frac{1}{2}\Delta\hat\rho 
-\nabla \cdot \big( \hat\rho(1-\hat\rho) \nabla H \big) 
+b(\hat\rho)e^H-d(\hat\rho)e^{-H} 
\end{equation}

As in Section~\ref{s:hj} we analyze the variational problem \eqref{quasipot}.
The associated Hamilton-Jacobi equation \cite{BJ} is 
\begin{equation}
\label{HJg+k}
\mf H \Big(\rho, \frac {\delta V}{\delta \rho}\Big)=0
\end{equation}
where the ``Hamiltonian'' $\mf H$ is not anymore quadratic in the momenta
and it is given by 
\begin{equation}
 \label{HA}
{\mf H}(\rho, H) =
\frac {1}{2} \langle H,  \Delta \rho \rangle 
+ \frac {1}{2} \langle \nabla H,\rho(1-\rho)  \nabla H \rangle 
- \big\langle b(\rho), 1-e^{H} \big\rangle   
 - \big\langle d(\rho), 1-e^{-H} \big\rangle   
\end{equation}

If $b$ and $d$ are as in \eqref{giada2} it is easy to find
the solution $V$ of \eqref{HJg+k}, \cite{GJLV}.
Let $\bar\rho=A_1/(A_1+A_2)$, the unique root of $b(\rho)-d(\rho)=0$,
then
\begin{equation}\label{easy}
V(\rho) = \int_0^1\! du\:
\Big[\rho\,\log\frac\rho{\bar\rho}+(1-\rho)\log\frac{1-\rho}{1-\bar\rho}\Big]
\end{equation}
If the reaction rates $c_x(\eta)$ are of the form \eqref{eqn26}, then
the invariant measure is Bernoulli and \eqref{easy} follows.  On the
other hand, as shown in \cite{GJLV}, there are choices of the reaction
rates such that \eqref{eqn26} fails but \eqref{giada2} holds.  In this
cases \eqref{easy} still holds and we may say that reversibility is
restored at the macroscopic level or that time reversal invariance is
violated ``weakly" by the microscopic dynamics.

\bigskip
\noindent\emph{Correlation functions.}\ 
In Section \ref{s:mgf} we studied long range correlations for some
boundary driven (hence non equilibrium) conservative models.  Here we
consider equilibrium states for the Glauber + Kawasaki dynamics, which
is non conservative, and we study their macroscopic correlation
functions.  In particular we show that, if the microscopic dynamics
violates time reversal invariance ``strongly", that is \eqref{giada2}
does not hold, long range correlations do appear, \cite{BJ}.

Recall that, in order for the system to be reversible, the rates
$c_x(\eta)$ of the Glauber dynamics should be of the form
\eqref{eqn26}.  Their relationship with the coefficients $b(\rho)$ and
$d(\rho)$ in \eqref{eqn6} is given in \eqref{giada1}.

Equation \eqref{HJg+k} is a very complicated functional derivative equation which,
as in Section \ref{s:mgf}, can be solved by successive approximations 
by formal power series expansion in $\rho - \bar \rho$.
Here $\bar \rho$ is a constant stationary solution of \eqref{eqn6},
i.e.\ a root of $b(\rho)-d(\rho)=0$.

Let $C$ be the correlation function. It gives the second order approximation of the
quasi-potential as in \eqref{expV}.
Proceeding as in Section \ref{s:mgf}, see \cite{BJ} for further details, we get that 
$C$ solves
\begin{equation}
\frac{1}{2}\partial_u^2C(u,v)- (d_1 - b_1)C(u,v)
-\frac{1}{2}\,\bar{\rho}(1-\bar{\rho})\,\partial_u^2\delta(u-v)
+b_0\delta(u-v)=0
\label{ce}
\end{equation}
where 
$$
b_1=b'(\bar\rho)\,\,,\,\,\,\, 
d_1=d'(\bar\rho) \,\,,\,\,\,\,
b_0=b(\bar{\rho})=d(\bar{\rho})=d_0
$$

Notice that, if \eqref{giada2} holds, we get
\begin{equation}\label{cond}
\gamma:=b_0-\bar{\rho}(1-\bar{\rho})(d_1-b_1)=0.
\end{equation}
and in this case, recalling \eqref{easy}, we of course have
$C(u,v)=\bar\rho(1-\bar\rho)\,\delta(u-v)$.
Conversely, a solution of the form $C(u,v)=\alpha\delta(u-v)$, for some $\alpha\ge0$,
exists only if \eqref{cond} holds and therefore $\alpha=\bar\rho(1-\bar\rho)$.

The above considerations imply that long range correlations 
do appear whenever \eqref{cond} fails.
In this case we say that irreversibility persists at the macroscopic level.
As in Section \ref{s:mgf}, we introduce the off-diagonal covariance $B$ such that
$$
C(u,v)=\bar{\rho}(1-\bar{\rho})\delta(u-v)+B(u,v)
$$
and we get that $B$ solves
\begin{equation}
-\frac{1}{2}\partial_u^2B(u,v)+(d_1 - b_1)B(u,v)=
\gamma\delta(u-v)
\label{cb}
\end{equation}
where $\gamma$ is defined in \eqref{cond}.
Note that $d_1 - b_1$, being the second derivative of
the potential calculated in a minimum, is positive.
Let $R=(d_1-b_1-(1/2) \,\Delta)^{-1}$ be the resolvent of the Laplacian 
on the torus.  
Then the solution of \eqref{cb} is
\begin{equation}\label{giadasol}
B(u,v) = \gamma\, R(u,v)
\end{equation}
Since $R(u,v)\ge0$, we conclude that the correlation $B(u,v)$ has the
same sign as $\gamma$.

While $\gamma=0$ corresponds to the macroscopically reversible situation,
in general $\gamma$ may have either sign.
For instance, given $\alpha\in(-1,1)$,
take the flip rates given by 
$$
c_0(\eta)=\eta_0\Big(1-\alpha\frac{\eta_{-1}+\eta_1}2\Big)
+(1-\eta_0)\Big(1+\alpha\frac{\eta_{-1}+\eta_1}2\Big)
$$
for $\alpha >0$ presence of surrounding particles enhances the birth
rate and suppresses the death rate. We thus expect that the two point
correlation to be positive for $\alpha>0$ and negative for $\alpha<0$. 
We have $\gamma=\alpha(1-\alpha)(2-\alpha)^{-2}$ which shows
that this is indeed the case.

In \cite{DFL} it is shown that fluctuations from the hydrodynamical equation 
with standard Gaussian normalization converge, as $N\to\infty$,
to an Ornstein-Uhlenbeck process. 
The stationary correlations of this process agree, as they should, with 
\eqref{giadasol}.

\subsection{Boundary driven asymmetric exclusion process}
\label{s:mtasep}

We start from the representation of the invariant measure
for the boundary driven totally asymmetric exclusion process obtained
in \cite{DuSch} and  illustrated in Section~\ref{s:tasep}. 
We call $\pi^N$ the empirical
measure associated to the configuration $\eta$, $\gamma^N$ the empirical
measure associated to $\xi$ and $\mathcal{G}(\rho,f)$ the
joint rate functional
\begin{equation}
\nu_N\left(\pi^N\approx \rho,\gamma^N\approx f\right)\sim
e^{-N\mathcal{G}(\rho,f)}
\end{equation}
where $\nu_N$ is the measure \eqref{invtasep}.
From formula \eqref{margi}, using the contraction principle of large
deviations, we obtain directly
\begin{equation}
\mu_N\left(\pi^N\approx \rho\right)\sim e^{-NV(\rho)}
\end{equation}
where
\begin{equation}
\label{altern}
V(\rho)=\inf_{f}\mathcal{G}(\rho,f)
\end{equation}
This argument suggests a different representation, from the one
obtained in \cite{DLS3}, for the non local rate functional
$V(\rho)$. In particular, while in \cite{DLS3} $V(\rho)$ is obtained
either as an infimum or a supremum, depending on the values of the
chemical potentials $\lambda_0$ and $\lambda_1$, 
here we write $V(\rho)$ always as an infimum.

We construct explicitly this new representation in the special case
$\lambda_0=\lambda_1=0$. In this case the measure $\nu^N$ is uniform
on the set of complete configurations, defined by \eqref{complete},
and the joint rate
functional $\mathcal{G}$ is easily obtained as a restriction of the
one associated to the uniform measure over all configurations
$(\eta,\xi)\in X^{\Lambda_N}\times X^{\Lambda_N}$. We
define the set of \emph{complete profiles}
\begin{equation*}
\mathcal{C}:=\Big\{(\rho,f)\,:\: \, 
\int_0^u\!dv\,  \big[ \rho(v)+f(v) \big] \geq u ,\, 
\; u \in[0,1] ; \; \int_0^1\!dv \, \big[ \rho(v)+f(v) \big] =1  
\Big\}
\end{equation*}
Remember that $\rho$ and $f$ are density profiles for 
configurations of particles satisfying an exclusion rule
so that they take values in $[0,1]$.
Then in the special case $\lambda_0=\lambda_1=0$ we have
\begin{equation}
\mathcal{G}(\rho,f)=
\left\{
\begin{array}{lll}
{\displaystyle 
\int_0^1 \! du \, \big[ h(\rho)+h(f) \big] } & 
\text{if} & (\rho,f) \in \mathcal{C}\\
+\infty & & \text{otherwise} \\
\end{array}
\right.
\end{equation}
where $h(x)=x\log(2x)+(1-x)\log[2(1-x)]$. Note that we do not need
to add a normalization constant due to the fact that the constant
profiles $\left(\frac{1}{2},\frac{1}{2}\right)$ belong to
$\mathcal{C}$ and
$\mathcal{G}\left(\frac{1}{2},\frac{1}{2}\right)=0$. Using
\eqref{altern} we obtain the following variational representation for the
quasi-potential $V$,
\begin{equation}
\label{alternes} 
V(\rho)=\inf_{\left\{f\,:\,(\rho,f)\in \mathcal{C}\right\}}
\int_0^1\! du \,  \big[ h(\rho) + h(f) \big]
\end{equation}
that has to be compared with the one in \cite{DLS3}
\begin{equation}
\label{derridatasep} 
V(\rho)=\sup_{\left\{f\,:\, f(0)=1\,, f(1)=0\,\right\}}
\int_0^1\! du\, 
\big\{ \rho \log[\rho(1-f) ] 
+ (1-\rho) \log[(1-\rho)f]+\log 4\big\}
\end{equation}
where the supremum is over monotone functions. 

In \cite{DLS3} it is shown that the supremum in \eqref{derridatasep}
is obtained when $f=F'_{\rho}$ and
$$
F_{\rho}(u)=CE \Big( \int_0^u\! dv \, [ 1-\rho(v) ] \Big)
$$
where $CE$ means \emph{Concave Envelope}. 
We will show next that
$F'_{\rho}$ is also the minimizer of the problem \eqref{alternes}.
This is equivalent to prove that
\begin{equation}\label{eqinf}
\inf_{\left\{ F\,:\, (\rho,F')\in \mathcal{C}\right\}}
\int_0^1\! du\, h(F'(u))=\int_0^1\!du \, h(F'_{\rho}(u))
\end{equation}
Note that $F'$ has to be a density profile so that $F$ is
increasing. 
Moreover $F$ is defined up to an additive constant so we can 
choose $F(0)=0$. 
The condition $(\rho,F')\in \mathcal C$ easily reads as 
\begin{equation}\label{cond*}
F(0)=0\,\,,\quad
F(u)\geq \int_0^u \! dv \, [1-\rho(v)] 
\,\, \forall\,u\in [0,1] \,\,,\quad
F(1)= \int_0^1\! dv \, [1-\rho(v)]
\end{equation}
It is clear that if $F$ satisfies condition \eqref{cond*}, then also
$CE(F)$ satisfies \eqref{cond*}; or, equivalently, if
$(\rho,F^\prime)\in\mc C$ then also $(\rho,CE(F)^\prime)\in\mc C$.
Moreover the following elementary inequality holds due to the
convexity of $h$,
\begin{equation}
\label{ineqsempl} 
\int_a^b\! du \, h(F'(u)) 
\geq(b-a) h\Big(\frac{F(b)-F(a)}{b-a}\Big)
\end{equation}
which immediately implies
$$
\int_0^1\! du\, h\big(F'(u)\big)
\geq \int_0^1\! du\, 
h\big( [CE(F)]'(u)\big)
$$ 
We thus conclude that we can restrict the infimum \eqref{eqinf} over
the set of concave functions $F$ satisfying conditions \eqref{cond*}.
Still a direct application of \eqref{ineqsempl} imposes that the
minimizer has to be the smallest among them, that is $F_{\rho}$.

Using the above result we can finally prove the equivalence between
the two different representations \eqref{alternes} and
\eqref{derridatasep} of $V(\rho)$.  In order to prove it we just have
to show that, for any density profile $\rho$,
\begin{equation}
\label{equadav} \int_0^1\! 
du \, \big\{ 
\rho\log(1-F'_{\rho}) + (1-\rho) \log F'_{\rho} 
\big\}
= \int_0^1\! du \, \big\{(1-F'_{\rho})\log (1-F'_{\rho})+F'_{\rho}\log
F'_{\rho}\big\}
\end{equation}
The contributions on both sides in
\eqref{equadav} from the domain of integration where
$F'_{\rho}=1-\rho$ are clearly equal. Consider now a maximal
interval $[a,b]$ where $F'_{\rho}\neq 1-\rho$. Then on this interval we have
$$
F'_{\rho}(u)=\frac{1}{b-a} \int_a^b\! dv [ 1-\rho(v) ] 
\,, \quad u\in [a,b]
$$
From this fact, an easy computation shows that also 
the contributions on both sides of
\eqref{equadav} from the integrations over $[a,b]$ are equal.

\bigskip\bigskip\noindent 
\normalsize\textbf{Acknowledgments}\par
\smallskip 
\par\noindent

{\small 
We are grateful to A. Faggionato, G. Basile and J. Lebowitz for very
illuminating discussions and collaborations.  G. J.-L. would like to
thank the organizers of the program \emph{Principles of the dynamics
  of non equilibrium systems} held at the
Newton Institute in 2006, for the invitation to participate.
Thanks in particular to C. Godreche for suggesting to write this paper.
We acknowledge the support of PRIN MIUR 2004--028108
and 2004--015228.
}

\end{document}